\documentclass{aa}

\usepackage{graphicx}
\usepackage{txfonts}
\usepackage{hyperref}
\bibpunct{(}{)}{;}{a}{}{,}

\usepackage[nolist]{acronym}
\usepackage{url}
\usepackage{paralist}
\usepackage{amsmath}
\usepackage{xspace}

\usepackage{verbatim}

\graphicspath{{figs/}}

\newcommand{\cgs}{\mathrm{erg}\,\mathrm{s}^{-1}\,\mathrm{cm}^{-2}}

\acrodef{ABRIXAS}[\textit{ABRIXAS}]{\textit{A Broadband Imaging X-ray All-Sky Survey}}
\acrodef{ACIS}{Advanced CCD Imaging Spectrometer}
\acrodef{AGN}{Active Galactic Nuclei}
\acrodef{AIP}{Astrophysical Institute Potsdam}
\acrodef{APE}{All-purpose Parameter Environment}
\acrodef{APS}{Active Pixel Sensor}
\acrodef{ARF}{ancillary response file}
\acrodef{ART-XC}{Astronomical Roentgen Telescope -- X-ray Concentrator}
\acrodef{ASCA}[\textsl{ASCA}]{\textsl{Advanced Satellite for Cosmology and Astrophysics}}
\acrodef{ASIC}{Application-Specific Integrated Circuit}
\acrodef{ASM}{All-Sky Monitor}
\acrodef{ASTEROID}{Active current Switching TEchnique ReadOut In X-ray spectroscopy with DePFET}
\acrodef{Athena}[\textsl{Athena}]{Advanced Telescope for High ENergy Astrophysics}
\acrodef{CAMEX}{CMOS Analog MultiplEXer}
\acrodef{CCD}{Charge-Coupled Device}
\acrodef{CDFS}[CDF-S]{\textsl{Chandra} Deep Field--South}
\acrodef{CdTe}{Cadmium Telluride}
\acrodef{CERN}{Conseil Européen pour la Recherche Nucléaire}
\acrodef{CESR}{Centre d'Etude Spatiale des Rayonnements}
\acrodef{CfA}{Center for Astrophysics}
\acrodef{cgs}{centimetre gram second}
\acrodef{CMB}{Cosmic Microwave Background}
\acrodef{CTE}{charge transfer efficiency}
\acrodef{CTI}{charge transfer inefficiency}
\acrodef{CXB}{Cosmic X-ray Background}
\acrodef{DEPFET}{Depleted P-channel Field Effect Transistor}
\acrodef{Dec}{Declination}
\acrodef{DLR}{Deutsches Zentrum für Luft- und Raumfahrt}
\acrodef{DUO}[\textsl{DUO}]{\textsl{Dark Universe Observatory}}
\acrodef{ECAP}{Erlangen Centre for Astroparticle Physics}
\acrodef{EPIC}{European Photon Imaging Camera}
\acrodef{eRASS}{\acs{eROSITA} All-Sky Survey}
\acrodef{eROSITA}[eROSITA]{extended ROentgen Survey with an Imaging Telescope Array}
\acrodef{ESA}{European Space Agency}
\acrodef{FET}{Field Effect Transistor}
\acrodef{FFT}{Fast Fourier Transformation}
\acrodef{FITS}{flexible image transport system}
\acrodef{FOV}{field of view}
\acrodef{FSC}{Faint Source Catalog}
\acrodef{FWHM}{full width at half maximum}
\acrodef{GEANT4}{GEometry ANd Tracking}
\acrodef{GenX}[\textsl{Gen-X}]{\textsl{Generation-X}}
\acrodef{GRB}{Gamma-ray burst}
\acrodef{GRAVITAS}[\textsl{GRAVITAS}]{\textsl{General Relativistic Astrophysics VIa Timing And Spectroscopy}}
\acrodef{GRXE}{Galactic ridge X-ray emission}
\acrodef{GSFC}{Goddard Space Flight Center}
\acrodef{GSL}{GNU Scientific Library}
\acrodef{HEALPix}{Hierarchical Equal Area isoLatitude Pixelization}
\acrodef{HEAO1}[\textsl{HEAO-1}]{\textsl{High Energy Astronomical Observatory~1}}
\acrodef{HEAO2}[\textsl{HEAO-2}]{\textsl{High Energy Astronomical Observatory~2}}
\acrodef{HDU}{Header and Data Unit}
\acrodef{HEAdas}{High Energy Astronomy data analysis system}
\acrodef{HEASARC}{High Energy Astrophysics Science Archive Research Center}
\acrodef{HEAsoft}{High Energy Astronomy software}
\acrodef{HESS}[H.E.S.S.]{High Energy Stereoscopic System}
\acrodef{HEW}{Half Energy Width}
\acrodef{HIFI}{HIgh Framerate Imager}
\acrodef{HIFLUGCS}{HIghest X-ray FLUx Galaxy Cluster Sample}
\acrodef{HMXB}{High-Mass X-ray Binary}
\acrodef{HPD}{Half Power Diameter}
\acrodef{HTML}{HyperText Markup Language}
\acrodef{HTRS}{High Time Resolution Spectrometer}
\acrodef{HXI}{Hard X-ray Imager}
\acrodef{HST}{Hubble Space Telescope}
\acrodef{IAAT}{Institute for Astronomy and Astrophysics Tübingen}
\acrodef{ICM}{Intra-Cluster Medium}
\acrodef{IDL}[\texttt{IDL}]{Interactive Data Language}
\acrodef{IF}{Intermediate Flare}
\acrodef{INTEGRAL}[\textsl{INTEGRAL}]{\textsl{INTErnational Gamma-Ray Astrophysics Laboratory}}
\acrodef{IRAP}{Institut de Recherche en Astrophysique et Planétologie}
\acrodef{ISCO}{Innermost Stable Circular Orbit}
\acrodef{ISDC}{INTEGRAL Science Data Centre}
\acrodef{ISIS}[\texttt{ISIS}]{Interactive Spectral Interpretation System}
\acrodef{ISM}{Inter-Stellar Medium}
\acrodef{ISS}[\textsl{ISS}]{\textsl{International Space Station}}
\acrodef{IXO}[\textsl{IXO}]{\textsl{International X-ray Observatory}}
\acrodef{JAXA}{Japan Aerospace Exploration Agency}
\acrodef{JFET}{Junction gate Field Effect Transistor}
\acrodef{LAD}{Large Area Detector}
\acrodef{LMC}{Large Magellanic Cloud}
\acrodef{LMXB}{Low-Mass X-ray Binary}
\acrodef{LOFT}[\textsl{LOFT}]{\textsl{Large Observatory For x-ray Timing}}
\acrodef{MAGIC}{Major Atmospheric Gamma-Ray Imaging Cherenkov}
\acrodef{MCP}{Micro Channel Plate}
\acrodef{MicroX}[\textsl{Micro-X}]{\textsl{High-Resolution Microcalorimeter X-ray Imaging Rocket}}
\acrodef{MIP}{Moveable Instrument Platform}
\acrodef{MIRAX}{Monitor e Imageador de RAios-X}
\acrodef{MIT}{Massachusetts Institute of Technology}
\acrodef{MOS}{Metal Oxide Semiconductor}
\acrodef{MOSFET}{Metal Oxide Semiconductor Field Effect Transistor}
\acrodef{MPE}{Max-Planck-Institut f\"{u}r Extraterrestrische Physik}
\acrodef{MPG}{Max-Planck-Gesellschaft}
\acrodef{NASA}{National Aeronautics and Space Administration}
\acrodef{NFI}{Narrow Field Imager}
\acrodef{NRTA}{Near Real Time data Analysis}
\acrodef{NS}{Neutron Star}
\acrodef{NuStar}[\textsl{NuSTAR}]{\textsl{Nuclear Spectroscopic Telescope ARray}}
\acrodef{OGIP}{Office of Guest Investigator Programs}
\acrodef{OOT}{Out Of Time}
\acrodef{PHA}{pulse height amplitude}
\acrodef{PI}{pulse invariant}
\acrodef{PIL}{parameter interface library}
\acrodef{pnCCD}[pn-CCD]{p-n~junction Charge-Coupled Device}
\acrodef{PSD}{power spectral density}
\acrodef{PSF}{point spread function}
\acrodef{QPO}{Quasi-Periodic Oscillation}
\acrodef{RA}{Right Ascension}
\acrodef{RASS}{\acs{ROSAT} All-Sky Survey}
\acrodef{rms}{root mean square}
\acrodef{ROSAT}[\textsl{ROSAT}]{\textsl{ROentgen SATellite}}
\acrodef{ROSITA}{ROentgen Survey with an Imaging Telescope Array}
\acrodef{RMF}{redistribution matrix file}
\acrodef{RXTE}[\textsl{RXTE}]{\textsl{Rossi X-ray Timing Explorer}}
\acrodef{SAO}{Smithsonian Astrophysical Observatory}
\acrodef{SAS}{Scientific Analysis System}
\acrodef{SASS}{Science Analysis Software System}
\acrodef{SDDs}{silicon drift detectors}
\acrodef{SGR}{soft Gamma-ray repeater}
\acrodef{SIMPUT}{SIMulation inPUT}
\acrodef{SIXTE}{SImulation of X-ray TElescopes}
\acrodef{SMC}{Small Magellanic Cloud}
\acrodef{SN}{SuperNova}
\acrodef{SPO}{silicon pore optics}
\acrodef{SQUID}{Superconducting QUantum Interference Device}
\acrodef{SRG}[\textsl{SRG}]{\textsl{Spektrum-Roentgen-Gamma}}
\acrodef{SXI}{Soft X-ray Imager}
\acrodef{SXS}{Soft X-ray Spectrometer}
\acrodef{TES}{transition edge sensor}
\acrodef{TLE}{Two Line Element}
\acrodef{TRoPIC}{Third R\"{o}ntgen Photon Imaging Camera}
\acrodef{USA}{United States of America}
\acrodef{VELA}{VLSI ELectronics for Astronomy}
\acrodef{WCS}{World Coordinate System}
\acrodef{WD}{White Dwarf}
\acrodef{WFI}{Wide Field Imager}
\acrodef{WFM}{Wide Field Monitor}
\acrodef{WHIM}{warm and hot intergalactic medium}
\acrodef{XEUS}[\textsl{XEUS}]{\textsl{X-ray Evolving Universe Spectroscopy}}
\acrodef{XGS}{X-ray Grating Spectrometer}
\acrodef{XIFU}[X-IFU]{X-ray Integral Field Unit}
\acrodef{XIS}{X-Ray Imaging Spectrometer}
\acrodef{XLF}{X-ray Luminosity Function}
\acrodef{XML}{extensible markup language}
\acrodef{XMM}[\textsl{XMM-Newton}]{\textsl{X-ray Multi-Mirror Mission Newton}}
\acrodef{XMS}{X-ray microcalorimeter spectrometer}
\acrodef{XPOL}{X-ray POLarimeter}
\acrodef{XRS}{X-Ray Spectrometer}
\usepackage{color}
\usepackage[normalem]{ulem}

\begin{document} 

   \title{SIXTE:  a generic X-ray instrument simulation toolkit}

   \author{ Thomas~Dauser\inst{1}\thanks{thomas.dauser@fau.de} 
     \and Sebastian~Falkner\inst{1} 
     \and Maximilian~Lorenz\inst{1} 
     \and Christian~Kirsch\inst{1} 
     \and Philippe~Peille\inst{2} 
     \and Edoardo~Cucchetti\inst{3}
     \and Christian~Schmid\inst{1} 
     \and Thorsten~Brand\inst{1}
     \and Mirjam~Oertel\inst{1}
     \and Randall~Smith\inst{4}
     \and J\"orn~Wilms\inst{1}
    }

   \institute{Remeis Observatory \& ECAP, Universit\"at
     Erlangen-N\"urnberg, Sternwartstr.~7, 96049 Bamberg, Germany \and
     Centre National d'Etudes Spatiales, Centre Spatial de Toulouse,
     Toulouse Cedex 9, France \and 
     IRAP, Université de Toulouse, CNRS, CNES, UPS, (Toulouse), France \and  
     Harvard-Smithsonian Center for Astrophysics, 60 Garden St.,
     Cambridge, MA 02138, USA 
}

   \date{Received ????; accepted ????}

   \abstract{ We give an overview of the SImulation of X-ray TElescopes ({SIXTE)}
     software package, a generic, mission-independent Monte Carlo simulation
     toolkit for X-ray astronomical instrumentation. The package is based on
     a modular approach for the source definition, the description of the optics,
     and the detector type such that new missions can be easily implemented. The
     targets to be simulated are stored in a flexible input format called
     \acs{SIMPUT}. Based on this source definition, a sample of photons is
     produced and then propagated through the optics. In order to model the
     detection process, the software toolkit contains modules for various
     detector types, ranging from proportional counter and Si-based detectors,
     to more complex descriptions like \ac{TES} devices. The implementation of
     characteristic detector effects and a detailed modeling of the read-out
     process allow for representative simulations and therefore enable the
     analysis of characteristic features, such as for example pile-up, and their
     impact on observations.
  
     We present an overview of the implementation of SIXTE from the input source,
     the imaging, and the detection process, highlighting the modular approach
     taken by the SIXTE software package. In order to demonstrate the capabilities of the simulation
     software, we present a selection of representative applications, including
     the all-sky survey of \acs{eROSITA} and a study of pile-up effects
     comparing the currently operating \textsl{XMM-Newton} with the planned
     \acs{Athena}-\acs{WFI} instrument. A simulation of a galaxy cluster with
     the \acs{Athena}-\acs{XIFU} shows the capability of SIXTE to predict the
     expected performance of an observation for a complex source with a
     spatially varying spectrum and our current knowledge of the future
     instrument.
  }

\keywords{instrumentation: detectors, methods: numerical, telescopes,
  X-rays: general}

\maketitle
   
\section{Introduction}
\label{sec:introduction}

Simulations play an important role in the development of astronomical
instrumentation and  in the scientific analysis of observations with
existing facilities. They provide the possibility to investigate the dependence of the
scientific capabilities of the facility on various instrumental
effects and allow us to test and verify whether or not the science goals of
the studied future mission will be achieved by a given design choice.

In the area of X-ray astronomy a variety of existing simulation tools have been
developed to address the particular needs of individual X-ray instruments.
Examples include \texttt{marx} \citep{wise1997a} for \textit{Chandra}
\citep{weisskopf2002a}, \texttt{SciSim} \citep{gabriel2005a} for \acsu{XMM}
\citep{Jansen2001a}, and
\texttt{NuSim} \citep{madsen2011a,zoglauer2011a} for \acsu{NuStar}
\citep{harrison2010a}. Traditionally, however, these software packages were
restricted to the particular mission for which they were developed. The only
exception is \texttt{simx}\footnote{\url{http://hea-www.harvard.edu/simx/}},
which enables fast simulation of observations for multiple X-ray missions,
including \textsl{Athena}, \textsl{NuSTAR}, and \textsl{Hitomi}. Nevertheless, the latter
software package does not allow detailed investigation of
timing-related features and includes only a limited range of detector types and
detector effects. While the simulator package is very useful to get a fast
overview of the capabilities of an instrument, by design \texttt{simx} is not well suited for more detailed simulation studies of the scientific performance of an
instrument.

\acs{SIXTE} (\aclu{SIXTE}) is designed to fill this gap. It strongly focuses on
a detailed modeling of detector effects for any type of complex astrophysical
source. Similar to simulators such as \texttt{marx} or \texttt{SciSim} its
purpose is therefore to provide scientists with realistic and representative
simulations of future instruments for complex astrophysical sources, while also
supporting instrument development with a detailed modeling of the detector
characteristics. The major difference to the aforementioned simulators is that
new instruments can be easily added without changes to the software or detailed
knowledge of the internal structure. In addition to the larger set of included detector
types, the modularity ensures that any component can be easily
replaced by a more complex and detailed description in the software.

Being able to simulate detector characteristics therefore puts \acs{SIXTE}
simulations far beyond a simple ``fakeit'' using simply the \ac{ARF} and \ac{RMF} to
predict the flux in each spectral bin for a given source model. Besides the
missing detector effects, fakeit is not applicable for extended sources,
brighter sources which may create pile-up, or timing analysis of variable
sources. On the other hand, contrary to a full simulator like \texttt{marx},
\acs{SIXTE} does not predict the behavior of the optics, but relies on the
calibration data such as vignetting, \ac{PSF}, \ac{ARF}, and \ac{RMF} as an input. This
approach greatly eases the computational effort needed for the simulation,
while keeping the flexibility to use simulated or measured calibration data of
arbitrary complexity to describe the instrument under study.

\acs{SIXTE} is currently the official instrument end-to-end simulator of
\acsu{eROSITA}
\citep{predehlEROSITASRG2014,predehlEROSITAMatedSRG2018} aboard \ac{SRG}
\citep{pavlinskyARTXCSRGOverview2018} and the planned \acsu{Athena} satellite
\citep{barcons2017} for both the
\acsu{WFI} \citep[\aclu{WFI},][]{meidingerWideFieldImager2017} and the \acsu{XIFU}
\citep[\aclu{XIFU},][]{barretATHENAXrayIntegral2018} instruments. \ac{SIXTE} was also used for
many different assessment studies, including studies on Simbol-X
\citep{ferrando2005}, the IXO \citep[international X-ray observatory,][]{barcons2011a,schmid2010a,schmid2011a}, the
\acsu{LAD} \citep[\aclu{LAD},][]{zane2012a} and the \acsu{WFM}
\citep[\aclu{WFM},][]{brandt2012a} aboard \acsu{LOFT}
\citep[\aclu{LOFT},][]{feroci2012a,schmid2012b}, the proposed \textsl{Arcus} grating
spectrometer for the NASA medium-class explorer program
\citep{smithArcusExploringFormation2017}, and the coded mask instrument \ac{HXI}
of the \acs{MIRAX} experiment
\citep{grindlay2011a,penacchioniProtoMIRAXHardXray2017}.

This paper is structured as follows. First we present an overview of the
implementation (Sect.~\ref{sec:implementation}), giving detailed information and
assumptions on the single steps from the definition of the astrophysical source
to the final output event file. A broad selection of applications is presented
in Sect.~\ref{sec:results} and a summary of the most important aspects and
conclusions in Sect.~\ref{sec:summary}. More detailed information on certain
important aspects is given in the Appendix.

\section{Implementation}
\label{sec:implementation}

\subsection{Overview}
\label{sec:impl_overview}

The \ac{SIXTE} simulator uses a Monte Carlo frame-work based on individual
photons. Before any simulation is performed, the information on the
astrophysical sources is collected in input files in a generic
\ac{FITS}\footnote{\aclu{FITS}}-based format called \acs{SIMPUT}\footnote{\aclu{SIMPUT}}
(Sect.~\ref{sec:simput}). This source list is then used in combination
with the pointing information for the instrument to generate photons
(Sect.~\ref{sec:photon_sample}). These photons are then propagated
through a representation of the optics of the mission, resulting in a list
of impact times, positions, and energies on the detector
(Sect.~\ref{sec:instrument}). In the final step, these impacts are
input into the detector model, which may include detection-related post processing to
generate the final event list (Sect.~\ref{sec:detector_model}).

\begin{figure*}
  \includegraphics[width=\textwidth]{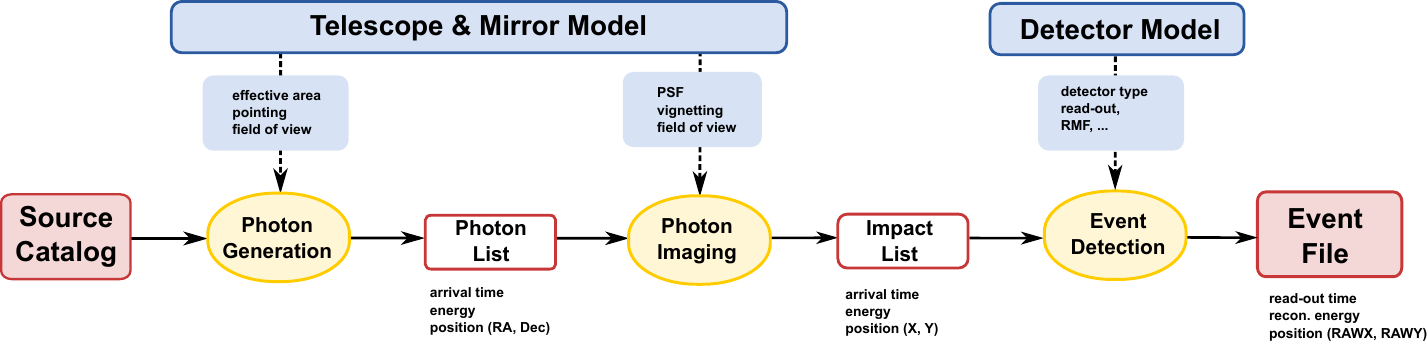}
  \caption{Flow chart illustrating the three major functional blocks of the
    \ac{SIXTE} simulation software (yellow boxes), starting from the
    generation of photons from a source catalog, which are then
    imaged and detected in the instrument and then processed to
    generate an event list. }
  \label{fig:sixte_pipeline}
\end{figure*}

Figure~\ref{fig:sixte_pipeline} highlights this modular approach
outlined above.  While the source definition does not require any
mission-dependent information, an increasing level of assumptions and
detector-specific details are necessary for the following steps. In
order to create a list of source photons, only knowledge of the
instrument effective area, the \ac{FOV}, and the pointing of the
instrument is required. Information on the latter two are crucial, as
a SIMPUT catalog will often contain sources in an area much larger
than the \ac{FOV} and ensure that only photons able to hit the
detector are generated. In order to process the photons through the
optics, additional information is necessary, such as the vignetting
and the \ac{PSF}. Only in the event-detection step does a detailed
description of the specific detector need to be given. Finally,
regardless of the detector type, a standardized event file is produced
as output of the simulation.

In the following sections the implementation of SIXTE is summarized.
Additional background information can be found in \citet{schmid2012a}
and the SIXTE simulator
manual\footnote{\url{https://www.sternwarte.uni-erlangen.de/research/sixte/}}.

\subsection{Source Description: \acs{SIMPUT}}
\label{sec:simput}

Performing simulations of realistic astrophysical objects requires
that the simulator be able to include any type of time variable or
extended source. For this purpose we developed the \ac{SIMPUT}
file format \citep{schmid2013a}, which provides an
instrument-independent definition of such sources. It is based on the
\acsu{FITS} data format
\citep[][]{wells1981a,ponz1994a,hanisch2001a,pence2010a}.

The main purpose of an instrument-independent source definition is
that the same input file can be used for simulations of different
instruments. Depending on the knowledge or assumptions on a particular
source, either basic or sophisticated models can be constructed with
phenomena ranging from simple energy spectra to spatial and temporal
variations of the observed spectra. 

The use of the \ac{SIMPUT} format is not restricted to \ac{SIXTE}, but
it is designed for general application. For instance, the simulation
software \texttt{simx} implements SIMPUT as an alternative input format.
In the following we give a brief summary of the basic characteristics
of the format. The full formal description of the format is given in
the relevant standard
document\footnote{\url{http://hea-www.harvard.edu/HEASARC/formats/simput-1.1.0.pdf}}.

\subsubsection{Source catalog}

\begin{figure}
  \centering
    \includegraphics[width=\columnwidth]{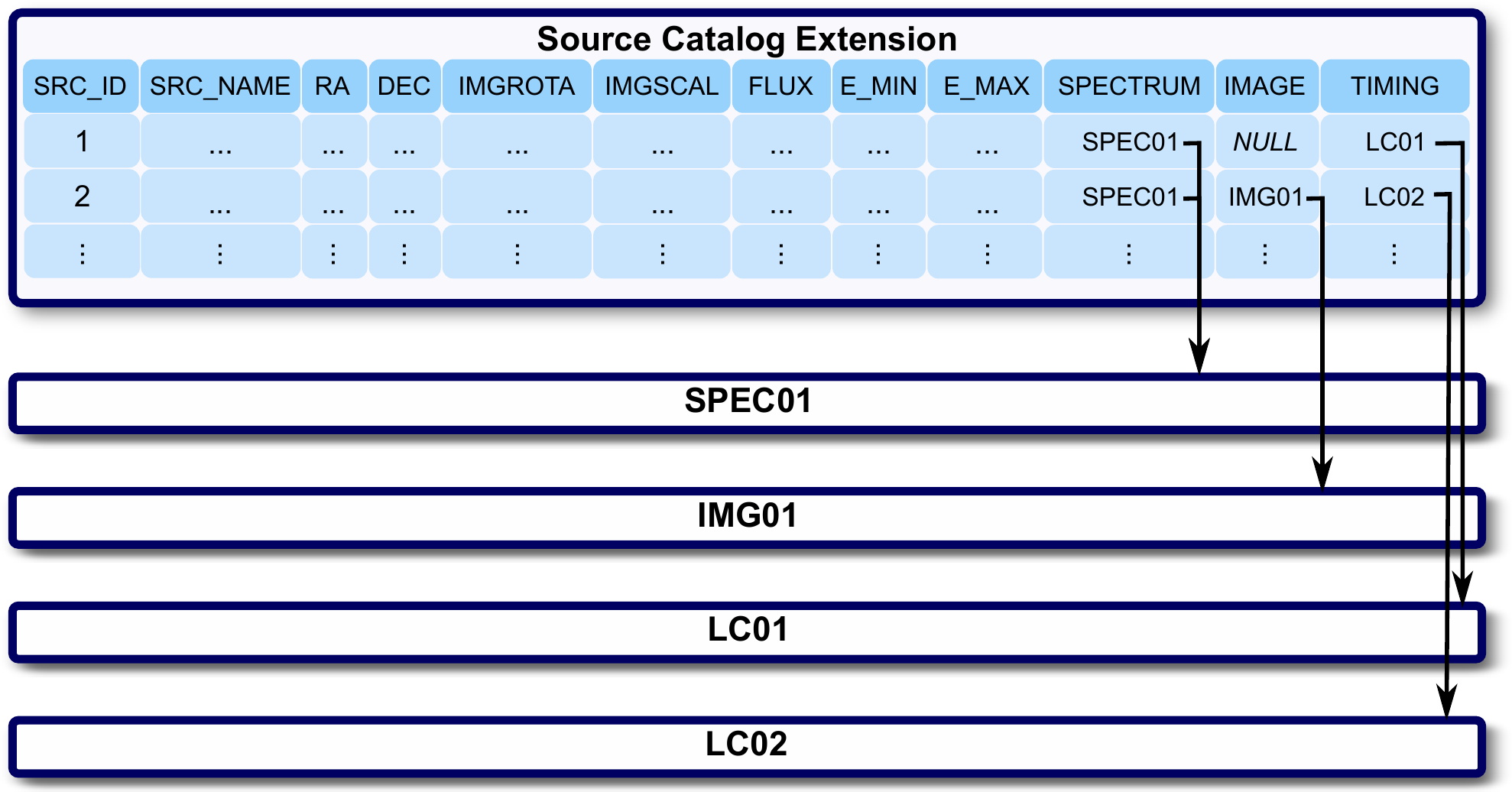}
    \caption[Schematic layout of a \acs{SIMPUT} file]{Schematic layout
      of a \ac{SIMPUT} file. The source catalog contains the main
      characteristics of one or multiple sources, such as their
      positions and observed fluxes. Additional data describing energy
      spectra, spatial extent, and time-variability are stored in
      separate \acp{HDU} and linked in the source catalog.}
  \label{fig:src_cat}
\end{figure}

The core of a \ac{SIMPUT} file is a source catalog specifying the properties of
one or multiple X-ray sources. The file consists of a table with entries for
each object defining its basic properties, such as its position and the observed
flux in a particular energy band. In addition, references to other \acsp{HDU}
(\aclp{HDU}) are given, which contain more detailed information about the energy
spectrum, time variability, and spatial extent of the source. These HDUs can
also be located in other SIMPUT files. While the reference to a spectrum is
obligatory for each source, the specification of time variability and spatial
extent are optional. The use of references to spectra and other source
characteristics allows their repeated use in large source catalogs. For example, when
simulating deep X-ray fields or an all sky survey with thousands to millions of
sources, spectra of individual sources from a certain class of objects can be
selected from a smaller set of source spectra which are representative of the
class as a whole. This approach allows the memory needed in the
simulation to be reduced. A schematic layout of a sample catalog is illustrated in
Fig.~\ref{fig:src_cat}.

\subsubsection{Energy spectrum}

The most common way to define a spectrum in \ac{SIMPUT} is the
specification of the photon flux density $F_\mathrm{p}(E)$
distribution in units of
$\mathrm{photons}\,\mathrm{s}^{-1}\,\mathrm{cm}^{-2}\,\mathrm{keV}^{-1}$.
Alternatively, the spectral shape can be modeled by a list of photons
stored in an \ac{HDU} table. The latter approach is convenient if the
source model is based on input from a hydrodynamic simulation,
which often provides a list of emitted photons.

As mentioned above, an individual spectral shape can be assigned to a
variety of sources with different brightness. Therefore the energy
spectrum given in the extension only describes the spectral shape,
which is then normalized to the flux of the source under consideration
which is determined by the reference flux assigned to this source in
the catalog \ac{HDU}.

\subsubsection{Time variability}

Time variability can be defined by a light curve, a \ac{PSD}, or a photon list
with time information assigned to each photon. A light curve describes the
time-dependence of the flux observed from a source. If the variability is
periodic, it can be defined with respect to the phase of the oscillation instead
of the absolute time. Such a description is quite suitable for modeling phenomena with a
typical time evolution such as type-I X-ray bursts or periodic signals from
pulsars. Power spectral densities{} are convenient to describe random and semi-random variations
such as noise or quasi periodic oscillations (QPOs). In order to obtain a proper
time distribution of simulated photons, they are internally converted to light
curves by the simulation software using the algorithm introduced by
\citet{timmer_koenig95a}.
Photon lists are often convenient to use if input is available from complex
three-dimensional simulations such as those used for example in cosmology. If the
number of  simulated photons approaches the total number of photons in the list,
the same input photons might be used multiple times.

\subsubsection{Source images}

Sources without an image assigned are considered point sources.
Extended sources are defined in the SIMPUT format by assigning an additional
\ac{FITS} image representing the spatial distribution of the observed flux or a
photon list that includes spatial information.
Both kinds of data can be obtained from an observation or a simulation of
the source. As for the spectrum, such images can also be used for multiple
sources as the reference point for the image, the total image flux, a scaling
parameter for the source extent, and a parameter describing the image
orientation are set in the source catalog.

\subsection{Photon generation}
\label{sec:photon_sample}

Based on the definition in the \ac{SIMPUT} file, a sample of photons
is produced with the following three basic characteristics required
for the subsequent instrument simulation: the time, the energy, and
the direction of origin of each photon.

\subsubsection{Source selection}

A \ac{SIMPUT} catalog can cover a sky area which is much larger than the part of
the sky observed by the telescope. To save computational time, photons are only
generated from selected sources relevant for simulation given the current
pointing of the telescope. This pointing can either be fixed to one position in
the sky, or optionally given as attitude in the simulation, specifying for each
time step the \ac{RA}, \ac{Dec}, and roll angle of the telescope.

Photons are therefore created for sources which have an angular separation
from the optical axis that is less than a certain preset angle from the current
pointing of the telescope. For typical simulations, this angle is set to 150\%
of the \ac{FOV} of the instrument. This conservative criterion takes into account
that simulations might involve a motion of the optical axis, for example, due to slews,
dithering, or pointing instabilities, and ensures that the statistical
properties of source light curves are not interrupted should the source move into
and outside of the FOV within one simulation run. For simulations
where stray light might be important, a larger angle might be required, which is
set in the configuration file for a given mission.

For catalogs with a large number of X-ray sources, selecting the
sources to be considered in the simulation can be computationally
challenging. For instance, an all-sky catalog of {active galactic nuclei (AGNs)} down to a
flux limit of $10^{-15}\,\cgs$ contains several million sources.
Due to the large number of sources, an efficient general search scheme
for all sources within a maximum angular distance is needed. 

In SIXTE, we use a general search algorithm based on $k$-d
trees, a general data type to store data in a $k$-dimensional space
\citep{bentley1975a,friedman1977a}. For optimized $k$-d~trees, the
computational effort of a range search for all sources within a
certain distance from a particular location scales as
${\cal O}(\log N)$, where $N$ is the number of sources in the
catalog \citep{friedman1977a,preparata1985a}. \citet{moore1991a}
describes an algorithm based on quick sort \citep{hoare1962a} that
allows  an optimized $k$-d tree to be constructed from a catalog that
scales as ${\cal O}(N\log N)$. While this step is computationally more
intensive, it has only to be done once at the start of the simulation,
and can be precomputed if necessary. In our experience, this step is
negligible compared to other steps of the simulation. Specifically,
within SIXTE we construct a 3D tree based on unit vectors pointing at
the catalog searches. To find all sources within a given angular
distance, $\theta$, from the pointing direction, we then use the tree
to find all sources within a linear 3D distance of $d=\sin\theta$.
We find that this approach scales well also to very large source
catalogs and yields roughly constant search times. We note that other
approaches to solve this problem which are used in astronomy, such as
searching on some variant of HEALPix coordinates
\citep{gorski2005a,calabretta07a}, are well suited for describing
images, but are less well-suited for the storage of catalogs of sources which
might be nonuniformly distributed on the sky.

\subsubsection{Photon arrival time}
\label{sec:photon_time}

The average photon rate $R$ observed from a particular source is
determined by its energy flux $F_E$ in the reference band
$E_\mathrm{min}$ to $E_\mathrm{max}$ specified in the \ac{SIMPUT}
catalog, the shape of the photon spectrum, $P(E)$, and the on-axis
\ac{ARF} of the instrument,
\begin{equation}
  \label{equ:photon_rate}
  R=
  \frac{F_E}{\int\limits_{E_\mathrm{min}}^{E_\mathrm{max}}
    P(E)\,E\,\mathrm{d}E} \cdot \int\limits_0^\infty
  P(E)\,\mathrm{ARF}(E)\,\mathrm{d}E\quad\textsf{.}
\end{equation}
We model the photon generation as a Poisson process with rate $R$. The
number $N$ of generated photons within a certain time interval $T$
therefore obeys a Poisson distribution with the expected value
$R \cdot T$. Starting at a particular point in time, the intervals
$\left(\Delta t\right)_i$ between subsequent photons are then
exponentially distributed and calculated by inversion sampling from
\begin{equation}  \label{equ:deltat}
  \left(\Delta t\right)_i = -\frac{1}{R}\ln\left(u_i\right)\quad\textsf{,}
\end{equation}
where $u_i$ denotes uniformly distributed random numbers
$u_i \in\left[0,1\right)$. We note that as we include the \ac{ARF} in the photon generation, the 
resulting photon list is specific to the observation and is  therefore instrument dependent. 
This approach greatly speeds up the simulation.

\begin{figure}
  \centering \includegraphics[width=0.9\columnwidth]{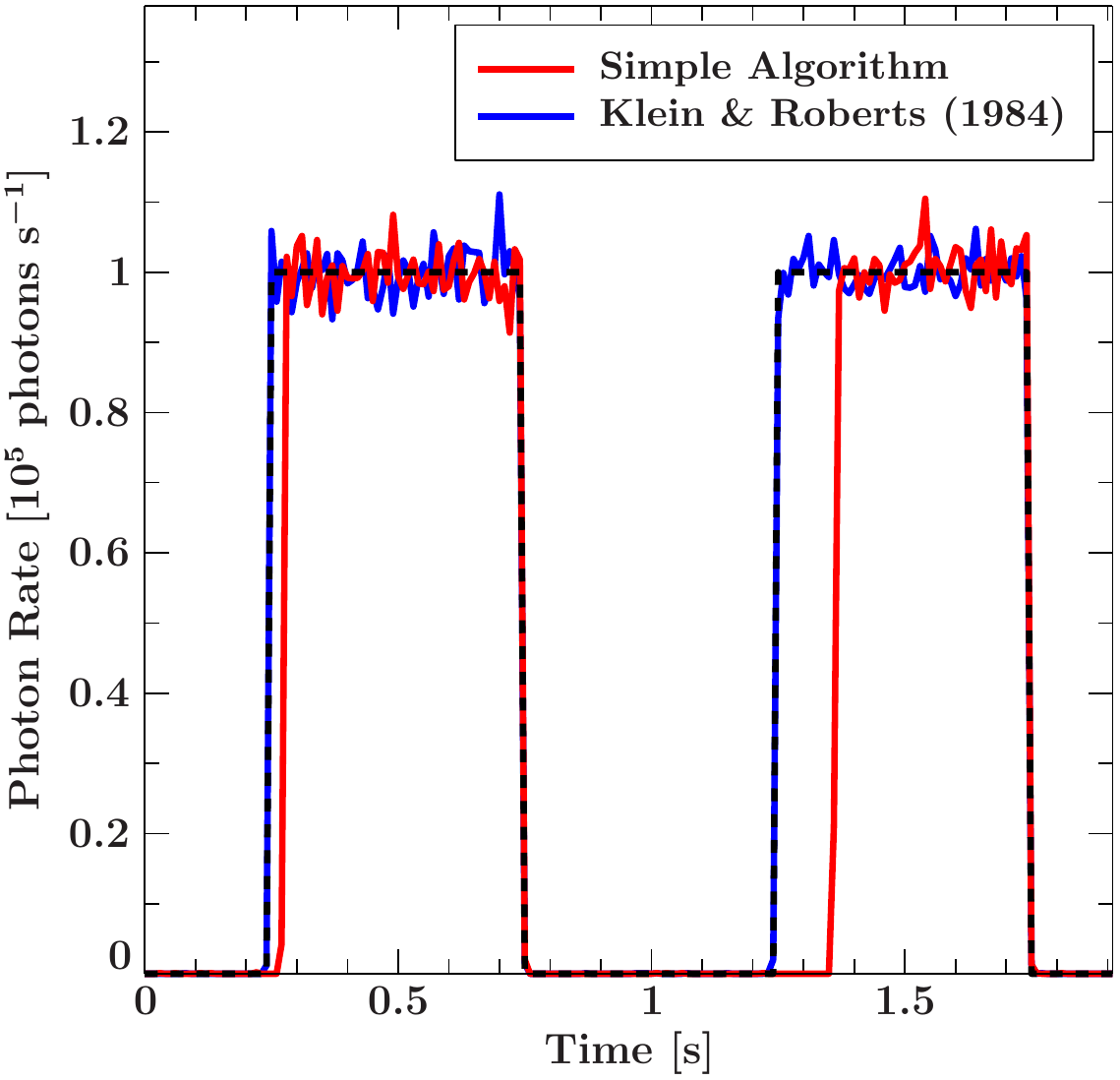}
  \caption[Poisson arrival process generator]{Comparison of different
    Poisson arrival process generators. The dashed black curve
    represents the input light curve. The red curve displays the
    simulated photon rate using a simple algorithm implementing
    Eq.~(\ref{equ:deltat}) with $R=r\left(t_{i-1}\right)$. The blue
    curve is based on the algorithm introduced by \citet{klein1984},
    which avoids the delay of the increasing photon rate for steeply
    rising profiles.} \label{fig:chopperwheel}
\end{figure}

While Eq.~(\ref{equ:deltat}) works well for sources with a constant
brightness, it is not appropriate for time-variable sources, even
though we have found quite a few implementations that do so. Although
the photon rate $R$ can be determined as a function of time, a fixed
value of $R$ is used for the evaluation of Eq.~(\ref{equ:deltat}). In
other words, the determination of the arrival time
$t_{i}=t_{i-1}+\left(\Delta t\right)_{i-1}$ of the $i$th photon
requires a particular value for $R$, which is usually chosen as
$R=r\left(t_{i-1}\right)$. This approach does not account for any
variations of $R$ within the interval $t_{i-1}$ to $t_i$. Particularly
for rapid changes from small to large values, Eq.~(\ref{equ:deltat})
can result in a delayed increase of the simulated photon rate, as
described by \citet{martin2009a} and illustrated in
Fig.~\ref{fig:chopperwheel}.

In \ac{SIMPUT} files the time-dependence of the rate function $r(t)$
is determined by the product of the light curve $l(t)$ of the
respective source, representing the time-dependence of the relative
flux variations, and the reference rate $R$, which is obtained
according to Eq.~(\ref{equ:photon_rate}) such that
\begin{equation}
  \label{equ:lc}
  r(t) = R\cdot l(t)
.\end{equation}
The light curve is approximated by a piece-wise linear function
\begin{equation}
  r(t) = a_n t + b_n \quad \mbox{for $s_n \leq t \leq s_{n+1}$}
,\end{equation}
with the supporting points $s_n$. We then use the Poisson arrival
process generator described by \citet{klein1984} to determine the
photon arrival times. This algorithm is basically an extension of
Eq.~(\ref{equ:deltat}) for piece-wise linear rate functions and avoids
the problems with steeply increasing photon rates.

\subsubsection{Photon energy}

The energies of the photons produced for a particular source are
distributed according to a spectral shape obtained by multiplication
of the instrument-independent spectral shape defined in the
\ac{SIMPUT} file with the instrument-specific energy-dependent on-axis
\ac{ARF} \citep{george1998a,george2007a}. Since the spectrum and the
\ac{ARF} are defined on a particular energy grid and for discrete
energy bins, respectively, an interpolation of the spectrum to the
energy bins of the \ac{ARF} is applied. Drawing random energies with
the inversion method (see \citealp{deak1990a}, p.~68ff.\ or
\citealp{gould2006a}, Ch.~11) reproduces the selected spectral
distribution.

\subsubsection{Photon direction of origin}
\label{sec:photon_direction}

For a point-like source, the direction of the incident photons is
equivalent to the location of the target on the sky and is trivially
included in the simulation. For a spatially extended source, we draw a
random photon direction based on the image defined in the
respective SIMPUT catalog. For this purpose the image is converted
to a cumulative distribution function,
\begin{equation}
  \label{equ:2d_pdensity} P(k,l) = \sum\limits_{i=0}^{k-1} \sum\limits_{j=0}^{M-1} p_{i,j} + \sum\limits_{j=0}^{l} p_{k,j}
,\end{equation}
with
\begin{equation}
  0\leq k<N \wedge 0\leq l<M 
,\end{equation}
where $N$ and $M$ are the width and the height of the image, and
$p_{i,j}$ is the normalized value of the pixel $i,j$ such that 
\begin{equation}
 P(N-1,M-1) = 1   \quad.
\end{equation} For each generated photon, a uniformly distributed random
number $r\in [0,1)$ is selected, and the pixel indices $k,l$
corresponding to the following relation, are determined in a
two-dimensional binary search,
\begin{equation}
\begin{aligned}
\label{equ:2d_binary_search}
  k &= \min\lbrace k^\prime \vert 0 \leq k^\prime < N \wedge P(k^\prime,M-1)>r \rbrace \\
  l &= \min\lbrace l^\prime \vert 0 \leq l^\prime < M \wedge P(k,l^\prime)>r \rbrace.
\end{aligned}
\end{equation}
For a sufficiently large number of simulated photons, this method
reproduces the input spatial flux distribution. Details on the
definition of the coordinate systems can be found in
Appendix~\ref{sec:coordinates}.

\subsection{Photon imaging: Optics implementation}
\label{sec:instrument}

In the photon imaging step, the photons are processed through the
optics and imaged onto the detector plane. The input consists of a
photon with an arrival time $t$, energy $E$, and the direction of its
origin in celestial coordinates, $\alpha$ and $\delta$. The outcome of
the photon projection step is photon impact with identical time $t$
and energy $E$, and a position $(X,Y)$ on the detector plane (see
Appendix~\ref{sec:detect_plane} for a definition of the coordinate
system used). If the photon does not hit the detector, it is
discarded.

In the current implementation \ac{SIXTE} uses pre-computed or measured
calibration files to describe this imaging process. While a separate ray-tracer
could be implemented, the chosen approach is computationally much faster and
allows the user to include measured calibration data. The reduction in detection
probability for off-axis photons compared to on-axis photons is described by a
vignetting function and the imaging properties by the PSF, both of which can
depend on energy, off-axis angle, $\theta$, and azimuth, $\phi$. For both
components, FITS standards exist \citep{george1994b,george1995a}. With
sufficiently detailed data derived from measurement or ray-tracing, this approach
mimics a full ray-trace almost entirely.  Point spread function asymmetries, spatially varying
vignetting, and even distortions due to single-reflection events can be
captured if desired. Conversely, a simple constant Gaussian PSF can be used for
missions in the early conceptual stage, giving this approach  great flexibility.

\begin{figure}
  \centering \includegraphics[width=\columnwidth]{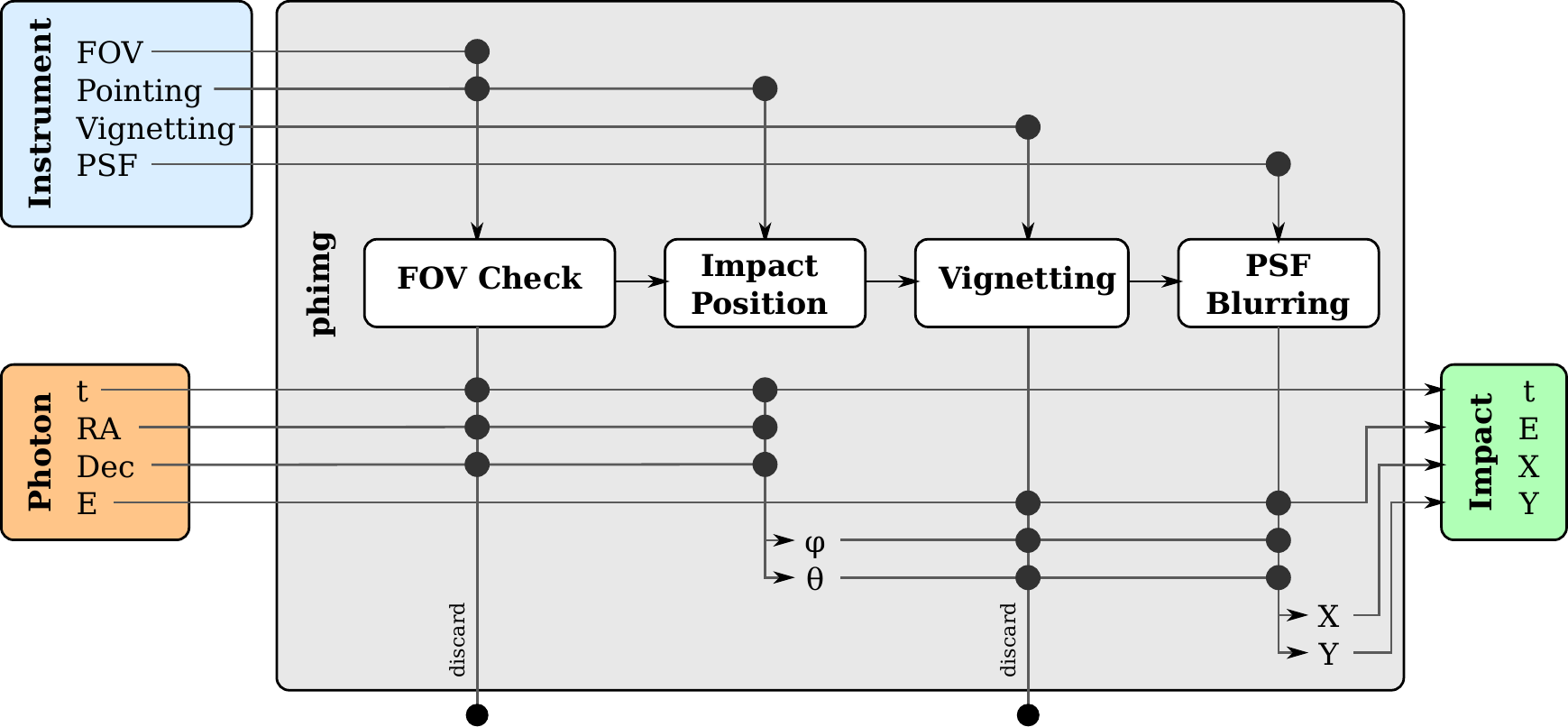}
  \caption{Block diagram of the photon imaging stage of SIXTE.}
  \label{fig:phimg}
\end{figure}

Figure~\ref{fig:phimg} shows an illustration of the flow of information in the imaging
stage of the simulation. The individual steps are covered
in the following.

\subsubsection{Vignetting correction}

The vignetting function describes the reduction in detection
probability for off-axis photons and has to be provided in the
standard calibration file format \citep{george1994b}. In the
simulation, vignetting is implemented as a discrete randomization
progress, based on the energy, off-axis, and azimuthal angle of each
photon. Depending on a probability distribution determined from these
parameters, the photon is either transmitted through the optical
system or is discarded.

\subsubsection{ Randomization of the PSF}

Focusing astronomical optics such as Wolter-type telescopes
\citep{wolter1952a} focus parallel rays of light onto the focal plane.
However, as the optics are not ideal, the image of a point source, the PSF, on
the detector is typically extended.  Most optics are designed to optimize the image
and thus the shape of the PSF inside a certain region around the
optical axis. With increasing off-axis angles, the PSF usually
degrades to a complex shape and grows larger in size \citep[see,
e.g.,][]{friedrichDevelopmentTestingEROSITA2012}.
 
To randomize the impact position of the photon on the detector, for most cases SIXTE uses a pre-calculated PSF as described in standard
FITS \ac{PSF} calibration files \citep{george1995a}. The PSF is
allowed to be dependent on the Energy, $E$, of the photon, the off-axis
angle, $\theta$, and the azimuthal angle, $\phi$, that is, the angle
between the intersection of the point $(X,Y)=(0,0)$ and the ideal
impact position and the $X$-axis of the instrument. The randomization is
calculated from the best matching entries in the calibration file,
weighted by the proximity of the incident position of the photon to these
entries. In order to avoid the computationally expensive 2D interpolation, this is implemented in \ac{SIXTE} by first
randomly drawing a single PSF --- respecting the calculated weight --- and
then drawing the impact position of the photon from the selected
PSF. \footnote{If complex types of 2D interpolation are
  desired, these have to be calculated by external tools and then
  provided to \ac{SIXTE} as a PSF file including the interpolated
  entries. }

In the case of azimuthal symmetry, the resulting PSF is optionally rotated
according to the difference between the simulated azimuthal angle and the one
of the PSF chosen from the calibration file. This reduces the blurring of some
asymmetrical features in complex PSF-shapes. This PSF is interpreted as a
spatial random distribution function for the final impact position of the
photon.

In case the instrument uses a collimator instead of imaging optics,
the impact positions of incident photons are randomly distributed on
the detector according to the geometry of the respective collimator
model. As for imaging optics, the off-axis transmission of the
collimator or coded mask is defined by a vignetting function.

Apart from simple collimators, hard X-rays are often imaged using
coded masks, which cast a particular shadow pattern on an imaging
detector, allowing the reconstruction of the location of the
illuminating sources \citep[][and references
therein]{cieslak16a,caroli87a,groeneveld1999}. A fraction of the
incident photons is absorbed by the opaque pixels of the mask. The
impact positions of the transmitted photons on the detector plane are
determined by geometrical considerations taking into account their
angle of incidence and the positions of randomly selected transparent
pixels.

\subsection{Event detection: Detector implementation}
\label{sec:detector_model}

\subsubsection{Overview}
After the photon generation and imaging described in the previous
section, the simulation has generated a list of the actual
\emph{impacts} in the detector plane. Due to fundamentally different
detector types, SIXTE provides several different tools to model the
detection process. The outcome of all of these steps is a standardized
list of events, the ``event file''.

Currently the \ac{SIXTE} simulator provides models for three different
types of detectors, which will be explained in the following:
\begin{itemize}
\item detectors working based on line-by-line read-outs, such as charged coupled device (CCD)
  or depleted field effect transistor (DEPFET) sensors (\textsl{eROSITA}, \textsl{XMM-Newton},
  \acsu{Athena} \acs{WFI}),
\item event-triggered detectors such as \ac{SDDs},
  proportional counters, or cadmium-zinc-telluride detectors
  (\textsl{Nustar}),
\item \ac{TES} calorimeter-type detectors (\ac{Athena} \ac{XIFU}), which
  are a special case of event-triggered detectors, where the energy
  resolution of each event depends on the time intervals between the
  previous and the following event.
\end{itemize}
Despite the different types of sensors, the simulation of event
generation follows a general structure. Mainly, this core
functionality of \ac{SIXTE} is broken down into two separate tasks:
(1) the signal creation in the detector and the signal read out, and (2)
the event reconstruction, and possible calibration. Between these two
steps a \emph{RawData} file is written to allow the user to
investigate the raw output or to perform a separate reconstruction and
calibration of the events.

The formal process of the simulation ends with the event read out. However, for
many applications it is necessary to apply the reconstruction and eventual
calibration in a post-processing step such that the final event file contains an
estimate of the reconstructed photon energy. This process then allows for a
spectral analysis of the simulated data. We note that for a calorimeter-type
detector such as the \ac{Athena}-\ac{XIFU} it is necessary to apply the
reconstruction and notably the event grading during the simulation, as it influences
the energy resolution of the events \citep[see][]{Peille2016}.

\subsubsection{Event creation}

In order to convert the photon energy to a signal in the detector, \ac{SIXTE}
applies a randomization of the signal based on the \acs{RMF} \citep[\aclu{RMF},
][]{popp2000a,edgar2011a}. This matrix describes the probability that a certain
signal is measured for a given input photon energy. Such a description is a
common element of the spectral analysis of X-ray data
\citep{george1998a,george2007a}, and allows for a fast simulation of the detection
process, while maintaining the characteristics of the specific detector and
absorbing material. Most importantly, this \ac{RMF} encodes the energy
resolution of the detector, but might also include additional detector effects
like escape peaks \citep[see, e.g., ][for more
details]{eggertSpectralResponseSilicon2006}.

In Si-type detectors, like CCDs or DEPFETs, the photon impact creates a charge
cloud at the incident location, which is simulated in the generic approach by a
2D Gaussian \citep{kimmel2006a,martin2009a}. For impacts close to the
edge of a pixel, this electron cloud can partly drift to the adjacent pixels and
create so-called split events. Details on the split model and the charge cloud
size are presented in Appendix~\ref{sec:charge_cloud}.

For the \ac{TES}-calorimeter, thermal or electrical coupling between pixels can
lead to a similar effect called \emph{crosstalk}. Due to this coupling, the
primary impact can induce a small signal in neighboring pixels. In case of
thermal crosstalk, this effect can be simply described in the instrument
configuration, by specifying the strength of this coupling depending on the
physical distance between pixels \citep[][, Kirsch et al., in prep.]{peille2018}

For all other types of detectors, the absorption of the photon is assumed to 
take place in the impact location, that is, fully confined to the pixel hit
by the photon.

\subsection{Event read-out}

The modeling of the event read-out is an important part of the overall
simulation. It assigns the time $t_\mathrm{e}$ and the position (RAWX/RAWY or
PIXID) to the final event. Depending on the type of read-out, this does not
necessarily coincide with the arrival time and impact location.

In order to define the detector, SIXTE provides the possibility to set up a
rectangular array of pixels with dimensions defined in an \ac{XML} file (see
Appendix~\ref{sec:xml} for an example). The pixel can then be uniquely
identified by the RAWX, RAWY coordinate. For special detectors such as the
hexagon-shaped \ac{XIFU} of \ac{Athena}, such a simple identification is not
possible. Instead, a unique ID (the so-called PIXID) is assigned to each pixel.

In the so-called \emph{time-triggered mode}, the signals generated by
incident photons are stored in the affected pixels and read out after
a specific exposure time. The corresponding sequence of detector
operations is defined in the configuration file. This method is used,
for example, for CCD or DEPFET detectors
\citep{boyle1970a,amelio1970a}. In such cases, the frame time of the
read out determines the time stamp of the event, i.e.,
$t_\mathrm{e}=t_\mathrm{frame}$.

Similarly, the location (RAWX/RAWY or PIXID) assigned to the event is determined
by the row in which the signal is detected. Depending on the read out type, this
position might not coincide with the actual impact position. For silicon-based
X-ray detectors, two distinctly different models are implemented in \ac{SIXTE}
which cover the majority of the currently operating X-ray telescopes.

In case of a CCD, the charge is shifted to the read-out anodes at the edge of
the detector. Photons absorbed during this charge shift will be registered with
incorrect spatial information \citep{lumb1991a} and create so-called out-of-time (OOT)
events. Charge loss due to charge transfer inefficiencies (CTIs), where charge is
lost while it is shifted over the detector, can also be taken into account.

In a DEPFET detector, the charge is not shifted, but read out
directly \citep[see, e.g., ][]{treberspurgStudiesPrototypeDEPFET2017}.
In such detectors, however, the detection process is more complex as
the digitization of the signal from the photon happens in the pixel
itself. This read-out process is modeled in the end-to-end simulator
in a three-step process. First the voltage signal of the photon together
with the signal baseline is measured. Then the charge is removed by a
clear signal and in a second integration step the baseline is
determined. In a third step, the measured signal is then determined by
taking the difference between the two integrations. In case a photon
hits the detector during the read-out process, the measurement will be
corrupted, creating a \emph{misfit} event. Depending on when exactly
the photon is absorbed in the read-out process, its energy is either
measured partially (first integration), split between the current and
the next frame, or registered with a negative energy (second
integration) and therefore discarded (see Appendix~\ref{sec:depfet}
for more information).

In the \emph{event-triggered} or \emph{continuous read-out mode}, each absorbed
photon is immediately registered as an event
$t_\mathrm{e}=t_\mathrm{photon}$. This mode is typical for
SDDs \citep{gatti1984a} or proportional
counters, and also for the special case of \ac{TES} calorimeters.
Optionally, a dead time after an event can be specified, during which
the detector is insensitive to further incident photons.

In some detectors, most importantly TES-type detectors, the precision by which
the signal can be measured depends on the time difference to the previous and to
the following event (in the same pixel).\footnote{A study of this effect for the
X-IFU is presented in \citet{peille2018}.} Therefore, depending on these time
intervals, multiple \ac{RMF}s can be used in SIXTE to describe the dependence of
the energy resolution on these time intervals.
 
\subsubsection{Background}
Different possibilities are provided in SIXTE to account for detector
background. Events induced by interactions of cosmic-ray protons with
the camera housing can be inserted from a data set obtained with
\acs{GEANT4} \citep{geant42003a,allison2006a} simulations. We note that
this data set needs to be screened for particle tracks before its
usage in \ac{SIXTE}. For a particular exposure time, a Poisson
distributed number of these events is selected and the corresponding
signals are added to the detector pixels \citep{wille2011a}.
Alternatively, random background signals can be obtained from a
distribution defined in a \ac{PHA} background file
\citep{arnaudk2009a}, which is commonly available for many
instruments. This latter method, however, does not allow to model the
more complex effects that background has on event reconstruction.

We emphasize that the above description concentrates only on the background
events that are created from interactions with highly energetic particles in the
sensors. The X-ray background is as equally important as this particle background, but  does not require special treatment in
SIXTE, as it can be simulated by adding respective sources to the \ac{SIMPUT}
catalog for the photon generation. For instance, instead of describing the
diffuse cosmic X-ray background ({CXB)} in a background \ac{PHA} file, it can be modeled by a large
number of individual {AGNs} \citep{gilli2007a}. Depending on the sensitivity
of the simulated instrument, this model of the {CXB} can be partly resolved
into individual sources, as with a real instrument. For the convenience of the user
we provide an example of such a SIMPUT file for
download\footnote{\url{https://www.sternwarte.uni-erlangen.de/research/sixte/simput.php}},
which can be added to any simulation to model the {CXB}.

\subsubsection{Event reconstruction}
For simulations of imaging instruments, the detector coordinates
assigned to the events can be projected back onto the sky in order to
obtain an image of the observed region. The corresponding routine
takes into account the pointing attitude of the telescope, which in
general is a function of time. The use of the \texttt{WCSLIB} library
enables the selection of different projection types
\citep{calabretta2002a}.

If the detector model implements charge cloud splitting, the produced
pattern types (``grades'') can be identified with an appropriate
search algorithm. As with the measured data of real instruments, it
can happen that a pattern is incorrectly identified with a single photon,
although it has been created by multiple photons hitting the same or
neighboring pixels during one read-out cycle of the detector. This
phenomenon is called ``pile-up''. Its correct modeling and
elucidation is essential for understanding instrumental effects for
observations of very bright sources
\citep{popp2000a,davis2001b,martin2009a}.

Additionally, as in real instruments an energy threshold is applied to
each signal that is measured. The type and level of this threshold can
again be set directly in the \ac{XML} file. Such a threshold is
typically applied to remove thermal noise from the measured signal.
For detectors producing split events, the lower energy threshold is of
special importance, as it influences the reconstructed energy.
Specifically, if the fraction of charge lost to a neighboring pixel is
below the lower energy threshold, it will not be detected, and therefore
it will not be taken into account during the pattern recombination. This will
yield a reconstructed energy smaller than the incident photon
energy and will influence the detector response, as discussed in the following section.

\subsubsection{Energy calibration}

Effects such as the lower energy threshold will lead to reconstructed photon
energies that are different\footnote{different in a sense that
  the distribution is not described by the \ac{RMF}} from the energy
of the incident photon. The energy can also be influenced by applying a nonzero
\ac{CTI}, or by a photon hitting the pixel of a {DEPFET}-type detector while
it is read out (more details in Appendix~\ref{sec:depfet}). Therefore, the
measured values of the signal (the so-called \ac{PHA}) should
never be seen as the reconstructed photon energy, but as the channel of the
instrument response function in which the photon is measured. The proper
assignment of the energy is only done in the reconstruction step.

It is clearly desirable for a simulator to explicitly predict such
effects. When trying to analyze these simulated data however, any
effect on the reconstructed energy needs to calibrated out. As the
simulator predicts these effects, the task of calibration does not
constitute a fundamental problem. It can also be used to produce
calibration files, which mainly consist of adding these corrections to
the \ac{RMF}, which can then be used in standard X-ray data analysis
tools (such as \texttt{Xspec} or \texttt{ISIS}) to analyze the
simulated data. We emphasize that there is a danger of double counting
when using a response matrix that already includes the effects of
charge splitting and the lower energy threshold for modeling the
detector itself.

If applicable and specified in the \ac{XML} file, in SIXTE such a
calibration is performed automatically. We note that in many cases it
might be desirable to investigate the uncalibrated events and
therefore this calibration is added as additional information in the
output event file as a \ac{PI} value instead of the
commonly used \ac{PHA}.

\subsection{The software}
\label{sec:software}

The functionality of the simulation code is contained in the two separate
software packages \ac{SIMPUT} and \ac{SIXTE}. All software is freely
available\footnote{\url{http://www.sternwarte.uni-erlangen.de/research/sixte/}}
and licensed under the GNU General Public
License\footnote{\url{http://www.gnu.org/licenses/gpl-3.0}}. The software runs
under Linux and Mac OSX and does not have any special dependences on other
software besides standard libraries such as \texttt{libgsl} or \texttt{libboost}.

The \ac{SIMPUT} package comprises the \ac{SIMPUT} library, which
contains basic functions to access \ac{SIMPUT} files, a set of tools
to handle the generation and management of these files, and functions
to produce photons for the sources specified through SIMPUT files, and
therefore can be used as a starting point in the development
of more specialized instrument simulations. As shown in
Sect.~\ref{sec:simput_tools}, these tools allow, for example, the easy
inclusion of an \texttt{Xspec} \citep{Arnaud1996a} spectral model in a
\ac{SIMPUT} file.

The \ac{SIXTE} package implements the setup and handling of instrument
models, building on the \ac{SIMPUT} library for the photon-generation
process. SIXTE provides the simulation tools for a generic detector
model and for multiple specific instrument models. An example of the
usage of these tools is shown in Sect.~\ref{sec:sixte_tools}.

The majority of the code is implemented in \texttt{C} and can be
compiled and installed using the GNU Autotools. Data are stored in
\ac{FITS} files using the \texttt{cfitsio} library \citep{pence1999a}.
In particular, as also mentioned in the previous sections, the
instrument calibration files follow the standards of
\acsu{HEASARC}\footnote{\url{http://heasarc.gsfc.nasa.gov/docs/heasarc/caldb/caldb_doc.html}}.
The interface of the implemented tools is designed similarly to the
\texttt{ftools}\footnote{\url{http://heasarc.nasa.gov/docs/software/ftools/}}
using the parameter interface library \citep[PIL;][]{borkowski2002a} or
\acsu{APE} to read program parameters and other functionality of the
\acsu{HEAsoft} package. Physical units in the \ac{FITS} files are
specified according to \citet{george1995c}.

The complete \ac{SIMPUT} and \ac{SIXTE} software package is managed by the
Remeis Observatory (Bamberg, Germany), including a helpdesk reachable by
email\footnote{\url{sixte-support@fau.lists.de}}. The software is managed under
GIT version control, and official versions are released every few months and
announced on the user mailing list\footnote{\url{sixte-users@fau.lists.de}}.
The correct functionality of each release is verified by an automated test suite before it
is published. Detailed information on the software including a large tutorial
can be found in the \emph{simulator manual}, which is constantly updated along
releases and can be downloaded from the homepage.

\section{SIXTE example studies}
\label{sec:results}

In this section we present examples of applications of
the \ac{SIXTE} software. Due to its generic implementation, a much
wider range of different instruments than the ones listed below can be
modeled. However, a comprehensive overview is not the intention of
this article. In the following we highlight different aspects and
use-cases for the simulator of existing and future detectors.

As \ac{SIXTE} is the official end-to-end simulator of the \acsu{eROSITA} mission
and both instruments on-board \textsl{Athena}, the WFI and the X-IFU, the
following examples focus on these missions to illustrate simultaneously the
wide range of applications and the reason for using the \ac{SIXTE}
simulator as a tool for science and detector simulations. The separated \ac{TES}
simulator included in the \ac{SIXTE} software has been described in
\citet{wilms2016}.

\subsection{All-sky survey with eROSITA}
\label{sec:erosim}

The \ac{eROSITA} experiment on \ac{SRG} consists of seven co-aligned
Wolter type~I telescopes, each with its own p--n junction {CCD (pnCCD)} detector
\citep{predehl2012a}. The main objective of the mission is the
performance of an all-sky survey in the X-ray band 0.5--10\,keV 
\citep[see][for more information]{merloni2012}.
The imaging properties of the telescopes
\citep{friedrich2008a,friedrich2012a} have to be well understood in
order to identify the large number of discovered objects as either
extended, such as galaxy clusters, or point-like, such as {AGNs}.
This task is especially challenging for weak sources since the number
of detected photons will be limited. In order to test and develop
algorithms for (transient) source detection before the start of the
mission, realistic simulations are crucial.

\begin{figure}
  \centering
  \includegraphics[width=\columnwidth]{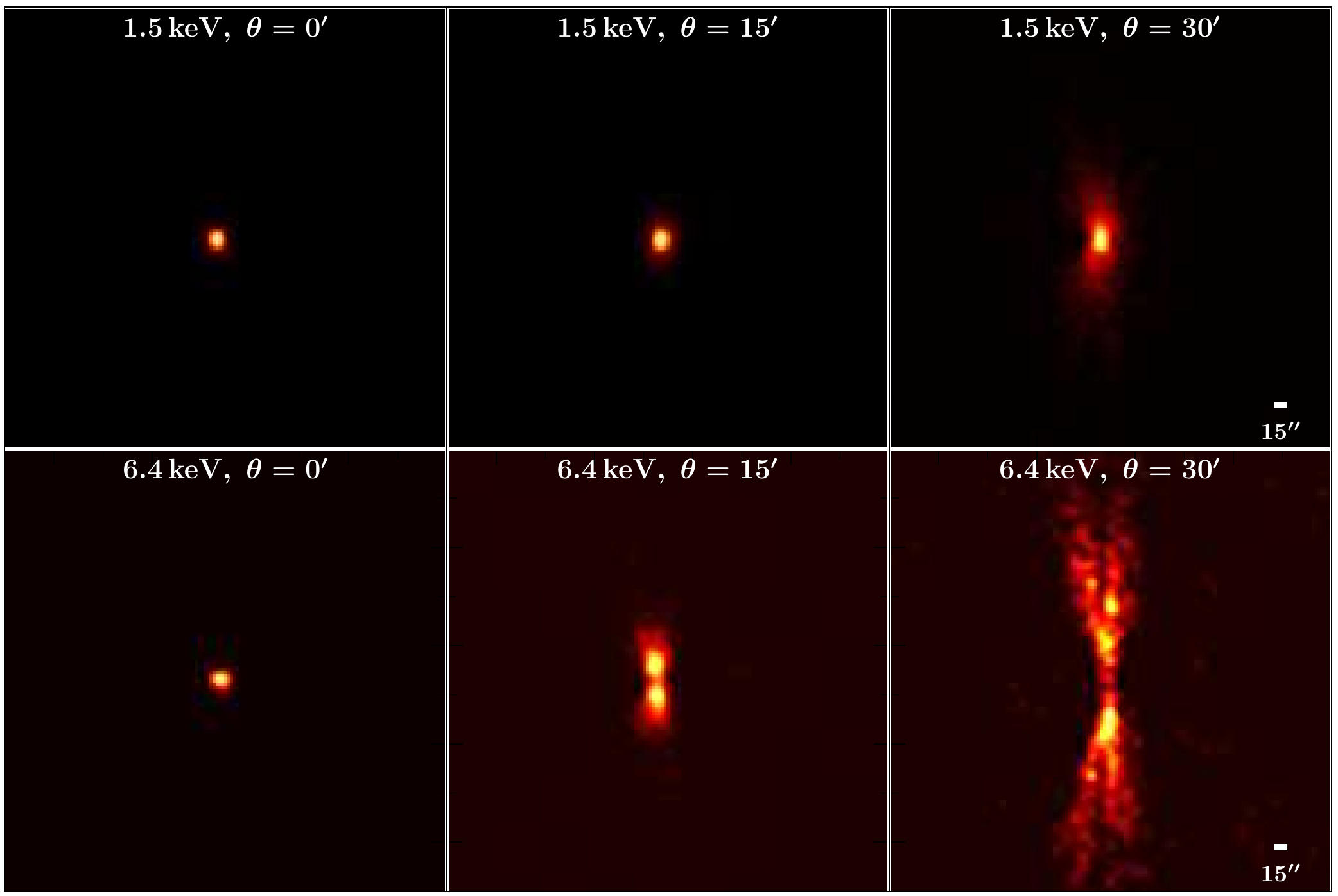}
  \caption{\ac{eROSITA} \ac{PSF} at 1.5\,keV for different off-axis
    angles from PANTER measurements
    \citep{burwitzCalibrationTestingEROSITA2014}. The on-axis \ac{PSF}
    has a \ac{FWHM} of $16\,\mathrm{arcsec}$, which deteriorates for
    larger off-axis angles leading to an average survey \ac{PSF} of
    $25$-$30\,\mathrm{arcsec}$ The images are in linear scaling, and
    are upscaled and smoothly interpolated from the measured data by a factor of
    three.}
  \label{fig:eropsf}
\end{figure}

\begin{figure}
  \centering \includegraphics[width=\columnwidth]{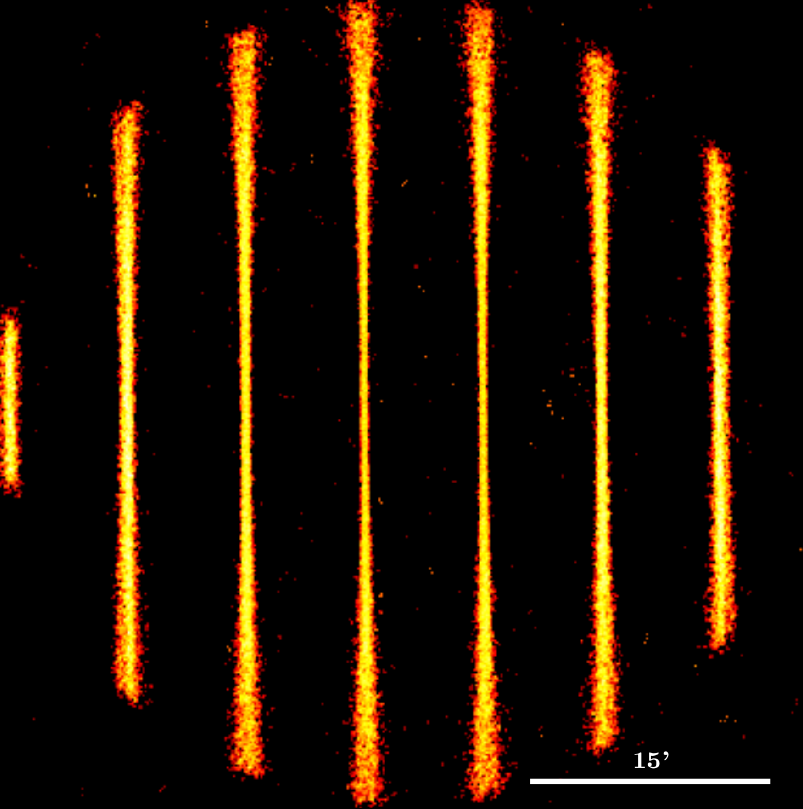}
  \caption{Detector image of multiple scanning observations over
    Cyg~X-1 during a half-year period of the \ac{eROSITA} all-sky
    survey. Due to the scanning motion of the spacecraft, individual
    passages over the source appear as stripes on the detector. Their
    width is relatively narrow in the center of the \ac{FOV} and gets wider
    towards the edge according to the different quality of the
    \ac{PSF}. In order to obtain an image of the observed target, the
    individual events have to be projected to the sky, taking into
    account the attitude of the telescope at the
    time of each observation. The image shows the complete \ac{FOV} of \ac{eROSITA}
    ($61\farcm2$ diameter) and uses a linear scaling.}
  \label{fig:cygx1}
\end{figure}

As shown in Fig.~\ref{fig:eropsf}, the \ac{PSF} strongly depends on
the off-axis angle of the incident photons. Since the \ac{eROSITA}
all-sky survey will be performed in a slew mode (illustrated in
Fig.~\ref{fig:cygx1}), the image of an individual source is determined
by a superposition of contributions from the \ac{PSF} at different
off-axis angles. The problem is even more complicated because the
alignment of the slew motion is different for different locations in
the sky, and the detectors are rotated with respect to each other.
Therefore, an accurate analysis of the imaging properties has to be
carried out for the position of the source in order to be
able to fully exploit the observed data.

\ac{SIXTE} is perfectly suited for this task because it can take into
account the motion of the instrument through the appropriate attitude
file as well as an energy- and angle-dependent model of the \ac{PSF}.
Importantly, the simulator allows the measured \ac{PSF}
and vignetting curve to be implemented into the simulation.

\begin{figure*}
  \includegraphics[width=\textwidth]{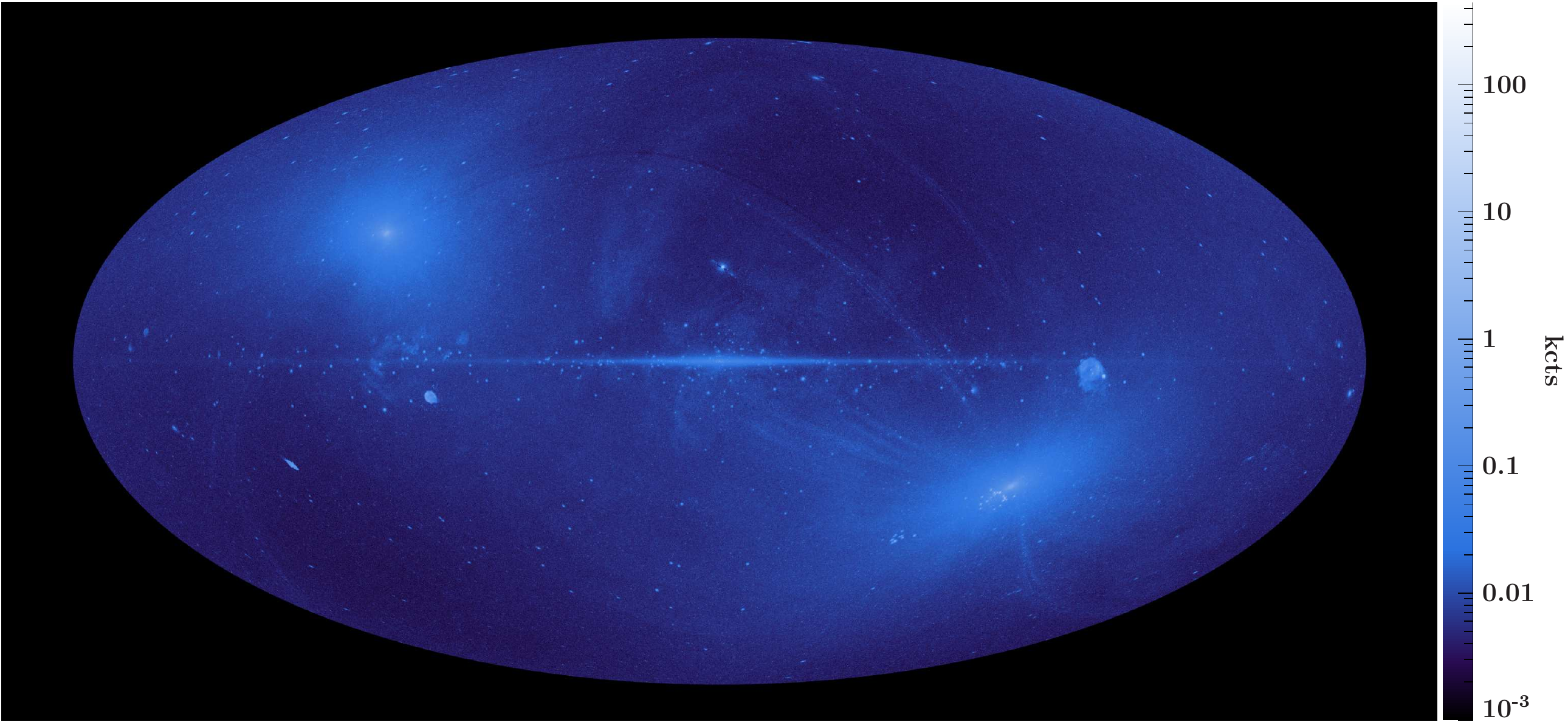}
  \caption[All-sky simulation of eROSITA]{All-sky simulation with the
    full eROSITA setup for a half-year survey. Details on the
    simulated sources can be found in the text.  Faint stripes are
    artifacts from the slightly inhomogeneous sampling of the extended
    ROSAT emission included in the simulation. We note that more
    counts are simulated at the poles, as the survey has a much larger
    exposure there and not due to differences in the sky brightness in
    these areas.}
  \label{fig:ero_allsky}
\end{figure*}

In the course of studying the overall survey performance of eROSITA, a
multitude of \ac{SIXTE} simulations have been performed. Whenever
possible, measured calibration data (RMF, ARF, PSF, vignetting) have
been used to create an output as close as possible to what eROSITA
will measure. One important example is a full all-sky simulation.

Figure~\ref{fig:ero_allsky} shows a representative half-year survey of eROSITA
simulated by \ac{SIXTE} using the predicted attitude (J.~Robrade, priv.\       
comm.). The input to the simulation consists of several different SIMPUT files,
including the complete \acs{RASS} \citep[\aclu{RASS},
][]{vogesROSATAllskySurvey1999}, the extended ROSAT soft X-ray background
\citep{snowden1997}, 589 variable bright sources from the \textsl{RXTE}-ASM, including
their light curves, the Galactic ridge emission from unresolved sources
\citep{turlerINTEGRALHardXray2010}, and extended objects such as M31 or the
Galactic center region.

This full all-sky survey of \ac{eROSITA} includes a complex instrument
setup containing measured calibration data, such as the PSF, combined
with the predicted attitude, and a simulation input consisting of many
different astrophysical objects. This simulation provided a rich data set for
pre-flight studies, highlighting its  capabilities and
also allowing for sensitivity studies, including a test of the
source detection software currently under development
\citep{brunnerEROSITAGroundOperations2018}. See \citet{clerc2018} for 
further examples of SIXTE simulations for \ac{eROSITA}.

\subsection{The \acs{CDFS} with \acs{Athena} WFI}
As a second example illustrating imaging simulations, we present the
result of a simulation of one snapshot of the \ac{CDFS} with the
\ac{WFI} large detector array, performed as part of a sensitivity
study for Athena. This region has been chosen as it is mostly
populated by very faint sources and has been subject to various long
observations with \textit{Chandra}
\citep{giacconi2002a,lehmer2005a,luo2008a,xue2011a} and \ac{XMM}
\citep{comastri2011a}. Therefore, rich input data exist allowing for
a realistic simulation and also providing the opportunity for a
detailed comparison with currently operating missions.

For the simulation, we used various types of input:
\begin{itemize}
\item For the central region, we converted the catalog provided
  by \citet{xue2011a} from multiple observations with \textit{Chandra}
  into a \ac{SIMPUT} file. Additional sources from the extended
  \ac{CDFS} \citep{lehmer2005a} complement the data in the area that
  is not covered by the previous catalog. The spectra of the sources
  in both catalogs have been modeled with an absorbed power law, which
  is a sufficient approximation, as the main aim of this simulation is
  a demonstration of the imaging capabilities of the instrument.
\item As the extended \ac{CDFS} still does not cover the entire
  \ac{FOV} of the \ac{WFI}, we added a sample of synthetic,
  randomly distributed AGNs to fill the gaps at the edge of the
  field. The source properties were chosen to resemble the observed
  $\log N$--$\log S$-distribution \citep[][and references
  therein]{hasinger2001a,brandt2005,cappelluti2007a,comastri2008a,cappelluti2009a}
  and the overall {CXB} spectrum following the distribution of
  spectral types according to \citet{gilli2007a}\footnote{We make use
    of the capabilities of the \ac{SIMPUT} file format to assign
    multiple sources to the same spectral model.}.
\item In addition, we compiled a catalog based on a list of galaxy
  groups \citep{finoguenov2014a} and a $\beta$-model
  \citep{king1972a,jones1999a} for the radial surface brightness
  profile of the galaxy clusters. The spectra are modeled with
  \texttt{apec} \citep{smithr2001a}.
\end{itemize}

\begin{figure*}
  \centering
 \includegraphics[width=\textwidth]{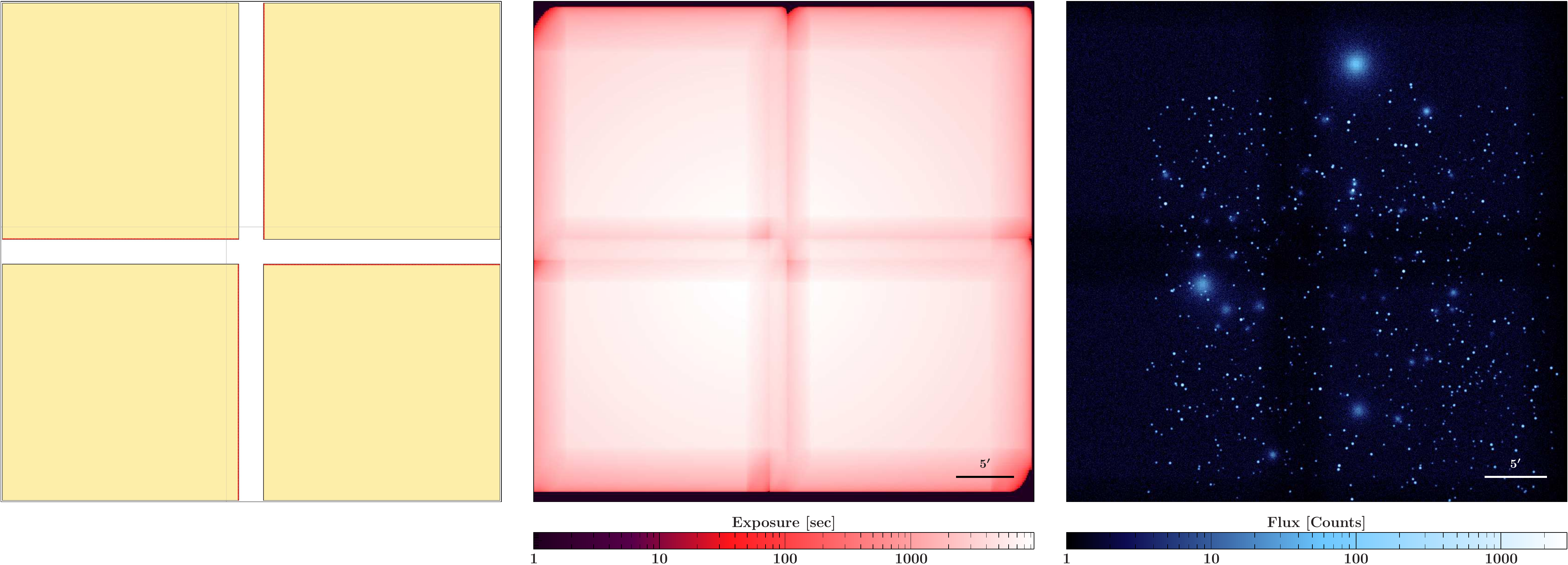}
 \caption{(a) Detector geometry of the WFI large detector array,
   showing the gaps between the chips and where the read-out is
   located (red line). (b) Exposure map showing the dithering used to
   more homogeneously distribute the exposure despite the gaps. (c)
   80\,ks observation of the \ac{CDFS} with the \ac{WFI} (40'
   \ac{FOV}), including dithering of the satellite.}
  \label{fig:wfi_cdfs}
\end{figure*}

The \ac{WFI} large detector array consists of four single DEPFET chips
(see Fig.~\ref{fig:wfi_cdfs}a), with a small gap between each one
\citep[see][]{meidingerWideFieldImager2017}. This detector setup
allows for a very large $40'$ \ac{FOV}, which, in combination with a
$5''$ PSF and the large effective area of \ac{Athena}, will allow the
detection of many very faint and distant objects in the early universe.
As sources situated in the gaps would therefore not be observed, it is
essential that \textsl{Athena} will be dithering during these observations.
\ac{SIXTE} easily allows the use of an attitude file to describe this dithering,
which we have chosen to be a Lissajous-type pattern. The exposure map is shown
in Fig.~\ref{fig:wfi_cdfs}b.

Finally, the simulated $80\,\mathrm{ks}$ observation of the \ac{CDFS}
is shown in Fig.~\ref{fig:wfi_cdfs}c, which evidently shows no gaps
due to the dithering over the large FOV. This simulation
illustrates the outstanding sensitivity of the \ac{Athena} \ac{WFI} in
just one snapshot, revealing a large number of distinct sources.
Further studies on this simulated data set can be performed, for example to
determine the limit for detecting faint sources.

\subsection{Merging galaxy clusters with the \acs{Athena} \acs{XIFU}}
\label{sec:cluster_xifu}

To illustrate a combination of high-resolution spectroscopy and imaging, we next
show a simulation that utilizes the X-IFU instrument on Athena. This detector
will be based on a large array of \ac{TES} operating at 50\,mK, which results
in an energy resolution of 2.5\,eV \citep{barretATHENAXrayIntegral2018}.
In combination with the good imaging capabilities of \ac{Athena} with a 5''
resolution over the 5' \ac{FOV} of the \acs{XIFU}, one of its main science goals
are studies of the properties of galaxy clusters through spatially resolved high-resolution spectroscopy.

As an example, here we show a single simulation of the starburst galaxy
M82 based on measured \textsl{Chandra} data. In order to account for
the change of the emitted spectrum over the cluster, we construct a
complex SIMPUT from the brightness distribution and two parameter maps
over the same region, giving the local absorbing column
($N_\mathrm{H}$) and the temperature of the collisionally ionized
plasma (modeled by the \texttt{apec} model). All data have been taken
and inferred from observations. Using the powerful
\texttt{simputmultispec} tool (see Appendix~\ref{sec:simput_tools}),
such a SIMPUT file can be automatically created from the input data.

\begin{figure*}
  \includegraphics[width=\textwidth]{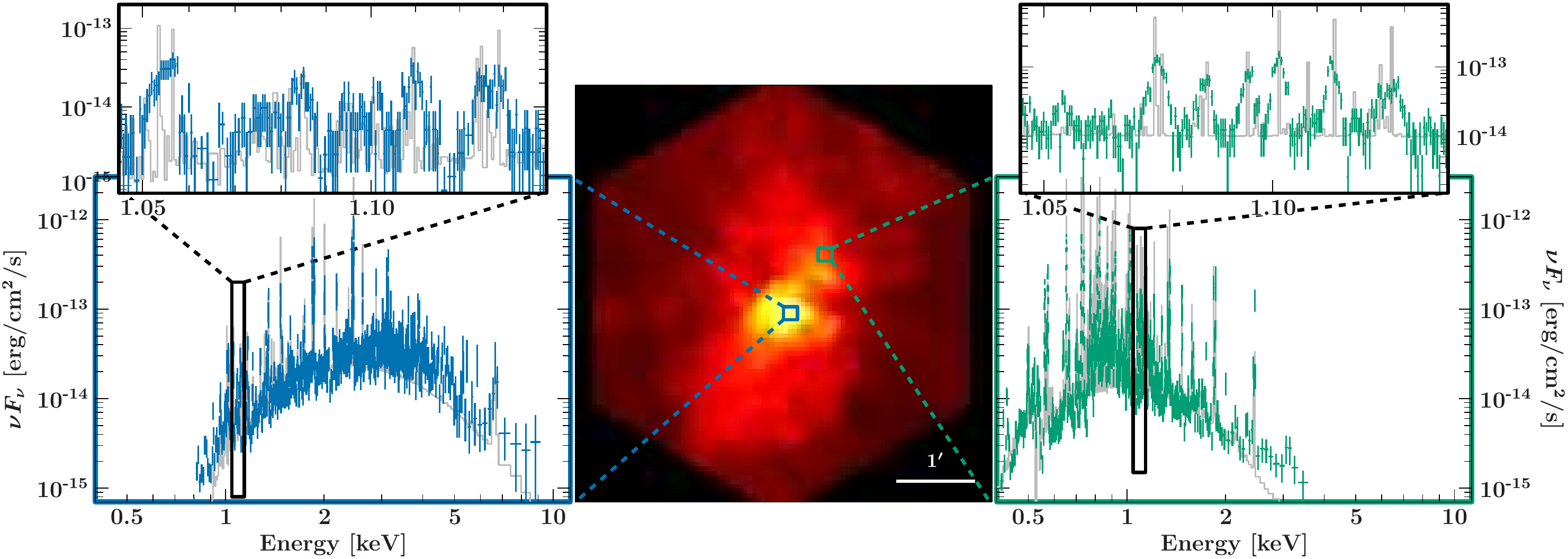}
  \caption{Simulated 100\,ksec observation of the starburst galaxy M82
    with the Athena X-IFU. The image in the center shows the inner
    arcminutes as simulated by \ac{SIXTE}. For the two regions (indicated
    by green and blue) we extracted spectra, which are plotted in the
    left and right panels together with the input model (gray line).
    For each spectrum we provide a zoom into the region around
    1.1\,keV to highlight the difference in observed lines. }
        \label{fig:m82}
\end{figure*}

Figure~\ref{fig:m82} shows the result of the simulations, with two spectra
extracted from the image to highlight that the \ac{XIFU} will be able to measure
detailed spectra, abundances, and line ratios over the extended source. We highlight the
superior signal-to-noise ratio and rich emission line spectrum, extracted from a
very small region of just $3\times3$ pixels out of more than 3000 total pixels for a
single 100\,ksec observation. Such simulations help to provide a
glimpse of the data we can expect from the X-IFU, importantly also allowing us
to develop methods for analyzing data sets of this kind, as \ac{Athena} will routinely
observe such sources.

Detailed \ac{SIXTE} simulations of galaxy clusters have been performed
by \citet{roncarelli2018} and \citet{cucchetti2018}. The latter study
is based on astrophysics simulations with the GADGET-3
smoothed-particle hydrodynamics code
\citep{rasiaCOOLCORECLUSTERS2015,biffi2018}. Using these SPH
simulations as input, the analysis of the \ac{SIXTE} simulations could
verify that the \ac{XIFU} is able to routinely measure the chemical
abundance and also the turbulent velocity in these clusters down to a
precision of 10\,km\,s$^{-1}$.

In \citet{brand2016} we performed another study,
which analyzes the capabilities of the \ac{XIFU} to detect the
\ac{WHIM} in absorption from measurements of Gamma-ray bursts (GRBs).
Such measurements of the baryonic matter in the universe are a core
science objective of the \ac{XIFU}, but also pose strong requirements
on how fast \textsl{Athena} has to react to alerts in order to be able
to detect such short events only lasting a few hours. Extensive
\ac{SIXTE} simulations showed that with the currently envisaged setup
\textsl{Athena} will be able to detect a few of these events per year.


\subsection{Simulating pile-up in silicon detectors} 
Besides complex science observations, \ac{SIXTE} is also very well
suited to supporting the detector development and characterizing the
detector performance. 

For brighter sources, one important issue in silicon-based detectors
is the increase of pile-up events in the detector. These events arise
when two photons hit the detector during the same read-out frame and
at the same position, and are therefore read out simultaneously. Such
events can not be separated and will be mistakenly identified as only
one event with the combined energy of both photons. For an increasing
source flux, the likelihood of such events increases and will
inevitably result in a change in spectral shape beyond a certain
threshold, possibly affecting the conclusions drawn from spectral
modeling. As a result, detectors will have a flux limit beyond which
the fraction of pile-up events is so large that spectroscopy becomes
impossible. As \ac{SIXTE} is able to correctly simulate the read-out,
the pile-up fraction can be predicted for any astrophysical source,
since, in contrast to real observations, we know which events are
piled-up in the simulation. Such simulations can be used to study the
performance of currently operating detectors (such as
\textsl{XMM-Newton}), or to predict the performance of observing
bright sources for future instruments. The latter is a very crucial
part of the instrument development, as will directly determine whether or not
the planned science goals will be achievable with the current
instrument setup.\footnote{In case of \textsl{Athena}, the study lead
  to the conclusion that a defocusing of the optics will be necessary
  to observe the bright objects.} We note that while there exist analytic
studies of the pile-up effect \citep[see][]{ballet1999}, simulations allow for much
more detailed and realistic estimations.

In order to study the effect of pile-up in silicon-based detectors, we simulate
a typical X-ray spectrum and scale it in flux. In order to ensure the same
statistical quality of each simulation, we adapt the exposure time such that
each obtained spectrum contains $\sim 10^6$\,counts. Importantly, the detailed
effect that pile-up has on the spectrum is strongly dependent on the spectral shape,
while its magnitude depends on the source flux. In the following we characterize
the flux in units of the Crab pulsar, which we set to be $F_\mathrm{crab} =
3.1\cdot10^{-8}\,\mathrm{erg}\,\mathrm{cm}^{-2}\,\mathrm{s}^{-1}$ in the 
0.2--10\,keV band
for a standard Crab spectrum of an absorbed power law with photon index
$\Gamma=2.1$ and equivalent hydrogen column $N_\mathrm{H}=0.4\cdot
10^{22}\,\mathrm{cm}^{-2}$. For a spectral shape deviating from this definition
we require the simulated detector to measure the same count rate as for this Crab
spectrum.\footnote{For example, with this definition a source with a flux of
1\,Crab in the \textsl{Athena}-WFI fast detector will therefore always produce
$\approx$74\,000 counts / sec regardless of its spectral shape.}

\begin{figure}
        \includegraphics[width=\columnwidth]{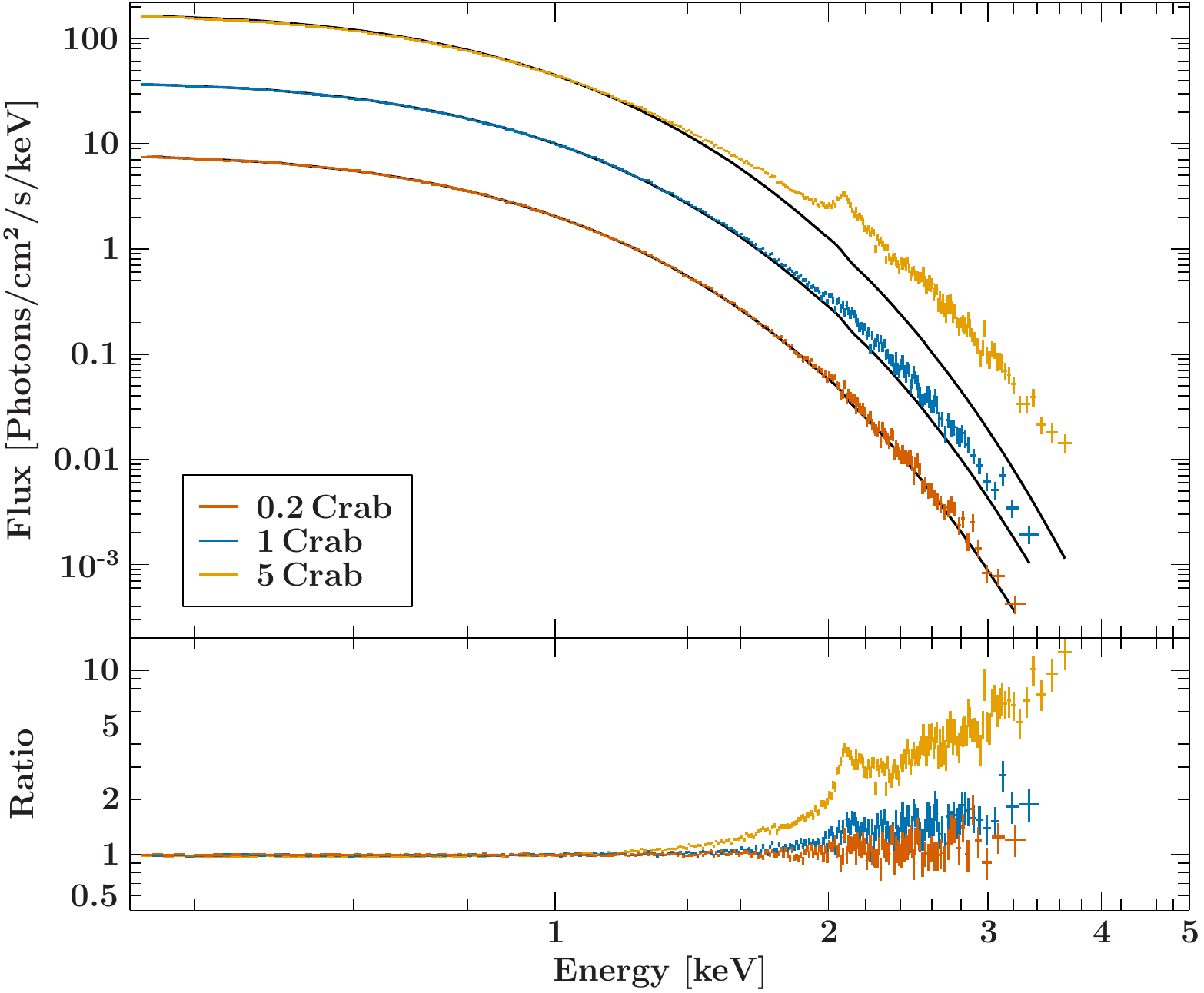}
        \caption{Spectra from an \textsl{Athena}-WFI fast detector simulation of a source with a 
        200\,eV black-body input spectrum. It can be seen that for increasing source
        flux the simulated spectrum starts to deviate from the input model (black line) at higher energies.}
        \label{fig:pileupsoft}
\end{figure}

To illustrate the effect pile-up has on the measured spectrum, we simulate a
very soft source with a black body with a temperature of 200\,eV.
Figure~\ref{fig:pileupsoft} shows a simulation of such a source with the fast
detector of the planned \textsl{Athena}-WFI instrument for an increasing flux of
0.2\,Crab, 1\,Crab, and 5\,Crab. These flux levels yield a fraction of piled-up
events among the total events of 0.15\%, 0.7\%, and 3.4\%, respectively. Since
pile-up means that two are seen as one event with the combined energy, there is
a systematic shift of flux to higher energies and therefore the spectrum
generally hardens with increasing pile-up.

In order to be able to quantify the effect of pile-up, we exemplarily 
use the above definition of the standard Crab as a typical power-law-shaped X-ray
spectrum. We note that a full study of pile-up for different kinds of spectral
models and source types is out of scope of this paper.

To compare the performance of a current instrument with that of a planned one, we
perform simulations for the EPIC~pn camera onboard \textsl{XMM-Newton} and for
the \textsl{Athena}-WFI. For each instrument we simulate three different
read-out modes, highlighting their different qualification to study bright
sources. While in the full-frame (ff) mode of the EPIC~pn one full chip is read
out, in the small window (sw) mode only $63\times64$ pixels are read-out. At the
cost of spatial information, the timing mode further improves the read-out
speed. It uses a $200\times64$ window and using line-shifts in the CCD it
combines ten~rows before the read-out\footnote{see
  \url{https://heasarc.gsfc.nasa.gov/docs/xmm/uhb/epicmode.html} for a detailed
  description}. For the \ac{WFI} we simulate a full-frame of
one large detector chip (large), a 64-row window mode (w64) of the same chip,
and the fast detector, which is the dedicated detector to observe bright sources
with \textsl{Athena}. The fast detector is implemented 35\,mm out of focus and consists of two
$32\times64$ pixel chips read out simultaneously.

\begin{figure*}
        \includegraphics[width=\textwidth]{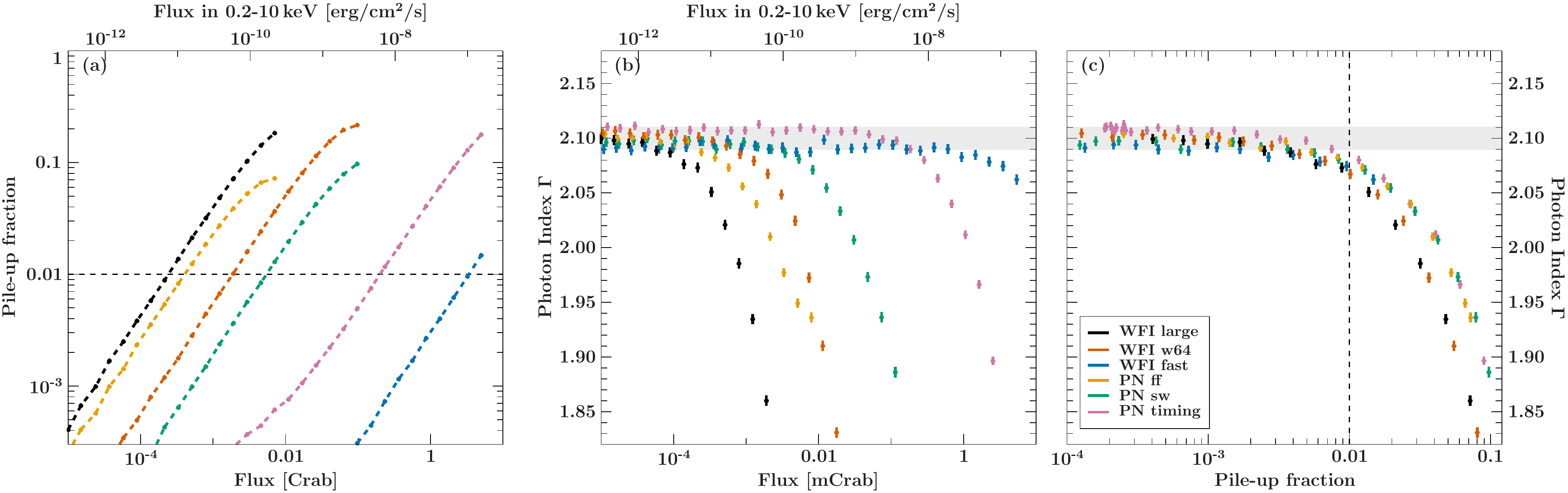}
        \caption{Simulation of a Crab-like spectrum with the EPIC~pn
          detector on \textsl{XMM-Newton} and the planned 
          \textsl{Athena}-WFI instrument for a selection of read-out
          modes with different qualification to study bright sources.
          (a) Fraction of piled up events in all detected events for
          increasing source flux. (b) Index of a power law fit to the
          simulated spectra, showing the effect of spectral distortion
          with increasing source flux depending on the detector and
          read-out mode. (c) Obtained power-law index plotted for
          the determined pile-up fraction, revealing a similar and
          generic behavior, largely independent of the detector. }
        \label{fig:pileup}
\end{figure*}

The results of the simulations are summarized in
Fig.~\ref{fig:pileup}. The most basic quantity is the so-called
pile-up fraction, which is the fraction of events affected by pile-up
out of all detected events. Figure~\ref{fig:pileup}a shows that as a
general trend this pile-up fraction is increasing with flux.
Clearly, the important difference is that the fast read-out modes
like the EPIC~pn timing mode or the WFI fast detector reach a
significant pile-up fraction of $\sim$1\% only at much higher fluxes.

For a deeper understanding of how these pile-up events affect the
spectral shape, we fit the simulated spectra with an absorbed power
law. The flux-dependent characterization of the spectral shape
(Fig.~\ref{fig:pileup}b) shows that for each read-out mode the
obtained power-law index at low count rates agrees very well with the
one of our input model ($\Gamma=2.1$). As expected, for increasing
flux there is a point for every mode where the determined $\Gamma$
starts to decrease, that is, the measured spectral shape hardens. This
result can be readily understood, because the energy of every pile-up event
is given by the sum of at least two source photons and we are
therefore artificially creating higher energetic events in the
detector.

Lastly, Fig.~\ref{fig:pileup}c, by plotting the photon index from
the
spectral fit as a function of the pile-up fraction of the respective simulated
observation, illustrates how the different instruments compare
in terms of their pile-up performance. The general behavior of all modes studied is similar. Most
importantly, for all modes studied, the spectral shape becomes significantly
distorted around 1\% pile-up fraction, indicating that observations should in
general strive to be below that fraction. We note that this limit only applies for
sources with a similar spectrum to our chosen input model of a power law with
$\Gamma=2.1$.

\section{Summary and Conclusions}
\label{sec:summary}

A detailed understanding of the instrumental features is required for
the development of new X-ray missions and for the analysis and
interpretation of data measured with existing instruments. This
knowledge can only partly be obtained from calibration measurements.
Simulations are required in order to analyze the complex phenomena
arising from the assembly of the various sophisticated components
making up the detection chain in an X-ray instrument. Specifically,
end-to-end simulations allow a realistic description of an
astrophysical source to be  connected with a solid modeling of the detection and
read-out process. In addition, simulated data allow comprehensive
verification tests of scientific analysis software even before the
launch of a new mission.

As demonstrated by the examples in Sect.~\ref{sec:results}, \ac{SIXTE}
provides many options for realistic simulations, covering various
different instrument technologies. The supported detectors range from
CCD detectors (\ac{eROSITA}, \textsl{XMM-Newton}), to more specialized
DEPFET detectors (\ac{Athena} \ac{WFI}), or a TES micro-calorimeter
(\ac{Athena} \ac{XIFU}). Additionally, the examples show that the
pointing of the instrument can be easily specified to include
dithering in an observation, or to simulate a full all-sky survey.

We also presented a study of the performance of a silicon-based detector
for observing bright sources. Both instruments, the currently
operating EPIC~pn onboard \ac{XMM} and the planned
\ac{Athena}-\ac{WFI}, show very similar behavior. As a general
result we conclude that above a pile-up fraction of $\sim$1\% the
spectrum becomes significantly distorted. The simulations also provide a
flux limit for each read-out mode, below which sources can be observed
without artificial spectral distortion from pile-up.

We have shown that the \ac{SIXTE} package constitutes a universal tool for
simulations of various X-ray instruments. The SIMPUT format allows
for any complex source input, including time variability with light curves or
power spectra, or extended sources with spectral variability. In order to ensure
a reasonable simulation time, the detection process is simplified by using
calibration files where possible (ARF, RMF), including a detailed representation
of the optics by allowing the vignetting and the PSF to change over the FOV. A detailed implementation is given for important characteristics such as
modeling the charge spread over the pixels, the detector read-out, or crosstalk
in TES devices. Depending on the accuracy of the input data and instrument
configuration, the simulations are therefore suitable to investigate effects
which are not accessible to analytical calculations or measurements, and to
connect scientific questions and goals with detector performance.

The software is designed to facilitate its use for scientific and
detector development studies. After defining the source input, the
simulation is run in a single step and produces an event file, which
can be readily analyzed with common X-ray data-analysis tools. It
already supports a large number of current and future X-ray
instruments, which can be easily adjusted to add new mission
configurations due to its modular concept and generic instrument model
defined in an XML-like configuration file.

\begin{acknowledgements}
  We thank Andy Ptak and Mihoko Yukita for providing the data used as
  input for the M82 simulation. The authors thank S.~Borgani,
  R.~Campana, M.~Ceballos, B.~Cobo, N.~Clerc, F.~F\"urst,
  P.~Friedrich, C.~Gro{\ss}berger, I.~Kreykenbohm, M. Martin, M.A.~Nowak,
  K.~Pottschmidt, M.~Roncarelli, F.-W.~Schwarm, P.~Weber and M.~Wille
  for helpful discussions and input. We thank the anonymous referee for valuable
  comments improving the paper. This work was funded by the
  Bundesministerium f\"ur Wirtschaft through Deutsches Zentrum f\"ur
  Luft- und Raumfahrt grants 50\,QR\,0801, 50\,QR\,0903, 50\,QR\,1103,
  50\,OO\,1111, 50\,QR\,1402, and 50\,QR\,1603. We thank John E.~Davis
  for the development of the \textsc{SLxfig} module used to prepare
  the figures. This research has made use of ISIS functions
  (ISISscripts) provided by ECAP/Remeis observatory and MIT
  (http://www.sternwarte.uni-erlangen.de/isis/) and quick-look results
  provided by the ASM/RXTE team. 
\end{acknowledgements}

\bibliographystyle{jwaabib}     
\bibliography{mnemonic,aa_abbrv,td_abbrv,detectors,additional} 

\appendix

\section{Coordinate systems}
\label{sec:coordinates}

\subsection{Sky coordinate system}

Unless specifically detailed, \ac{SIMPUT} and \ac{SIXTE} use an
equatorial coordinate system with RA and Dec to describe position on
the sky. The transformation between the sky images or coordinates and
the detector plane is described by the \ac{WCS}
\citep{greisen2002a,calabretta2002a} and performed with the
\texttt{WCSLIB}\footnote{\url{http://www.atnf.csiro.au/people/mcalabre/WCS/wcslib/}}
library. Any FITS images as source input will therefore be correctly
transformed, as long as they contain a valid \ac{WCS} header
describing the transformation from the given image coordinates to sky
coordinates.

\subsection{Instrument coordinate system}
\label{sec:detect_plane}
The instrument coordinate system $(X, Y)$ is defined as a planar
Cartesian system located in the focal plane behind the instrument
optics. Its units are in meters, with the origin being the point of
intersection between the optical axis and the focal plane. If an
attitude file is used which has the keyword \texttt{ALIGNMEN} set to
\texttt{MOTION}, the $X$-axis of the instrument coordinate system
points along the direction of motion of the pointing vector. In any
other case, the $X$-axis points towards the celestial north pole. The
$Y$-axis then points towards the west. If additionally a roll angle is
defined in the attitude file, the $X$- and $Y$-axes as described
before are rotated around the optical axis by this angle. The roll
angle is defined clockwise, as projected onto the sky.

\subsection{Detector coordinate system}
\begin{figure}
  \centering \includegraphics[width=\columnwidth]{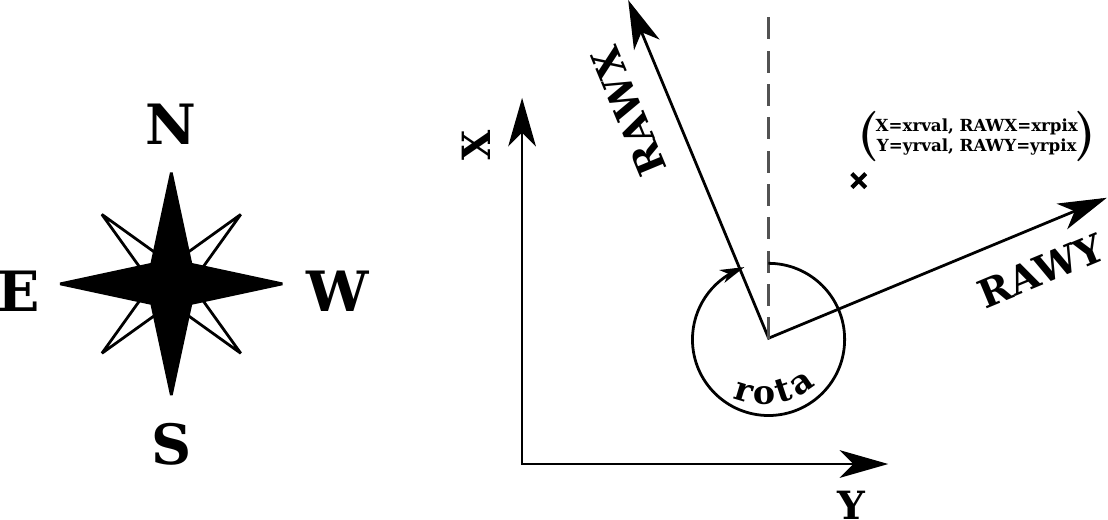}
  \caption{Coordinate systems as used by SIXTE, seen as projected
    onto the sky. In this figure, the X-axis points towards north. It
    can be realigned using an attitude file. The RAWX and RAWY axes
    can be rotated by the \texttt{rota}-key in the detector XML
    description. The angle is measured from the $X$- to the RAWX-axis
    clockwise when projected onto the sky.}
  \label{fig:coord_sixte}
\end{figure}

To describe a detector, the detector coordinate system is introduced.
Its new axes are named RAWX and RAWY, and are measured in units of
pixels. The physical size of the pixels is defined via the keys
\texttt{xdelt} and \texttt{ydelt}. We note that the coordinate system can
be shifted with respect to the instrument system. This is done via the
XML-detector description and its WCS coordinate tags. The keys
\texttt{xrval} and \texttt{yrval} define a point in the $(X,Y)$-plane.
The keys \texttt{xrpix} and \texttt{yrpix} define the respective pixel
coordinates which are associated with this point. The new system can
also be rotated via the \texttt{rota}-key using the angle from the X-
to the RAWX-axis, measured clockwise when projected onto the sky
(i.e., the same definition as the roll angle).
Figure~\ref{fig:coord_sixte} illustrates the coordinate systems as used
by SIXTE.

\section{Tools and ssage}

The \ac{SIMPUT} and \ac{SIXTE} software come with a large number of
tools tailored to different purposes and to facilitate the definition
of simulation input and to perform simulations. A complete overview is
given in a dedicated \emph{simulator manual}, which is constantly updated
along with the software and can be downloaded from the \ac{SIXTE}
homepage\footnote{\url{http://www.sternwarte.uni-erlangen.de/research/sixte/}}.

In the following a nonexhaustive list of important tools is given
to highlight the range of possible applications of the simulator.

\subsection{\acs{SIMPUT} Tools}
\label{sec:simput_tools}

As mentioned in Sect.~\ref{sec:implementation}, the \ac{SIMPUT}
software package contains a set of tools enabling the generation and
handling of \ac{SIMPUT} files based on the respective library
routines. The following list summarizes the available programs:

\begin{description}

\item[\texttt{simputfile}:] Generate a \ac{SIMPUT} file containing a
  catalog with a single source, a spectrum, and optionally a light
  curve or a \ac{PSD} and an image. A sample program call is listed
  below.

\item[\texttt{simputmerge}:] Merge multiple \ac{SIMPUT} files.

\item[\texttt{simputmultipsec}:] Create complex \ac{SIMPUT}-files
  where the spectrum changes over an extended region on the sky (e.g.,
  a galaxy cluster). As input, it needs a brightness distribution, a
  spectral model, and one or several parameter maps, parametrizing the
  spatial change of the given model. The tool will then automatically
  create a SIMPUT, encoding the spatial change of the model (see the
  \ac{XIFU} simulation of M82 in Sect.~\ref{sec:cluster_xifu}).
\end{description}
In order to illustrate how easily \ac{SIMPUT} files can be generated
with the appropriate tools of the \ac{SIMPUT} software package, we
show the assembly of a file with a single point-like X-ray source
located at $\mathrm{RA}=23.8^\circ$, $\mathrm{Dec}=-12.9^\circ$. The
spectral model for the source is provided in the file
\texttt{model.xcm}. With these input data, the file can be generated
by the following command:
\begin{verbatim}
simputfile Simput=source.simput \
   Src_ID=1 Src_Name=mysource \
   RA=23.8 Dec=-12.9 \
   XSPECFile=model.xcm Emin=2 Emax=10
\end{verbatim}
Of course, \ac{SIMPUT} files can also be constructed using the
library routines in other programs as well as by generating the
relevant FITS structure ``by hand'' using one of the many available
libraries generating FITS such as CFITSIO or pyFITS.

\subsection{\acs{SIXTE} tools}
\label{sec:sixte_tools}

The \ac{SIXTE} software package contains a set of tools to perform simulations
with different X-ray instruments. A selection of the most relevant tools is
given below.
\begin{description}

\item \texttt{runsixt}\\Performs a simulation using the generic
  instrument model. It performs the full simulation as detailed in
  Sect.~\ref{sec:photon_sample}-\ref{sec:detector_model}, from generating X-ray
  photons according to the source definition, imaging in a Wolter-type
  telescope, modeling the detection process, and performing the pattern
  recombination task. It outputs a standard event file. We note that there are separate 
  tools as well, to perform the individual tasks outlined above.

\item \texttt{xifupipeline}\\Similar to \texttt{runsixt}, but allows simulation
of calorimeter-type detectors such as the X-IFU. Allows for a more flexible
assembly of pixels (e.g., in a hexagon). The simulation includes the event
grading, which depends on the separation of the events in time, and optionally also 
thermal and electrical crosstalk effects.

\item \texttt{erosim}\\Derivative of \texttt{runsixt} specialized to
  the setup of the \ac{eROSITA} experiment, to allow the simultaneous simulation
  of all seven sub-instruments.

\item \texttt{athenawfisim}\\Derivative of \texttt{runsixt}
  specialized to the large detector array setup of the \ac{Athena} \ac{WFI},
  consisting of four chips.

\item \texttt{makespec}\\Produces a \ac{PHA} spectrum
  \citep{arnaudk2009a} from the event list generated by the detector
  simulation.

\item \texttt{makelc}\\Produces a light curve \citep{angelini1994a}
  from the event list generated by the detector simulation.

\item \texttt{ero\_vis}\\Determines the visibility of X-ray sources at
  particular positions in the sky during the \ac{eROSITA} all-sky
  survey based on the corresponding attitude file.

\item Coded mask detectors can be simulated in a threefold process \citep[see
][]{oertel2013a}:
 (1) \texttt{comaimg} models the photon absorption and transmission
  process through a coded mask and  determines their impact positions on the
  detector  (2) \texttt{comadet}  processes these impact positions and produces
  an event list \citep{oertel2013a}.
 (3) \texttt{comarecon}
  reconstructs the positions of the illuminating sources based on the shadow
  pattern on the detector, which is arising from the absorption by the coded
  mask. 

\end{description}

The following example of a simulation with a model of one of the seven
\ac{eROSITA} sub-instruments illustrates the usage of these tools:

\begin{verbatim}
runsixt \
   XMLFile=erosita.xml \
   Attitude=allskysurvey.att \
   Exposure=86400.0 \
   Simput=source.simput \
   EvtFile=events.fits \
\end{verbatim}

This program call initiates a simulation of a 1-day interval of the
\ac{eROSITA} all-sky survey based on the attitude specified in the
\ac{FITS} file \texttt{allskysurvey.att}. The model of the instrument
is defined in the \ac{XML} file \texttt{erosita.xml}. The observed
X-ray source is described in the \ac{SIMPUT} file
\texttt{source.simput}. The output of the simulation is stored in the
\ac{FITS} file \texttt{event.fits}.

\section{XML instrument configuration}
\label{sec:xml}

One of the key features of \ac{SIXTE} is the possibility to enable simulations
for a wide range of different X-ray instruments. The  necessary flexibility is
achieved by using a generic telescope and detector module, which can be
configured using a specific \ac{XML}-like format. This generic approach has
proven invaluable both for developing a wide coverage of various existing
instruments with different operation modes and for implementations of new
instrumentation in the course of several assessment studies. Especially for the
latter purpose the \ac{XML} format is very convenient for studying modifications
of an instrument setup.

For all supported instruments, including different, relevant read-out modes, the
\ac{XML} configuration files, including additional instrument calibration files,
are available for download on the \ac{SIXTE} homepage.

Below, we illustrate the power of this concept giving examples for the setup of
an \ac{eROSITA} sub-instrument, the burst mode of the
\ac{EPIC}~pn aboard \ac{XMM}, and the burst option of the \ac{XIS}
aboard \textit{Suzaku}. More details on this format, such as a description of
the XML tags, are given in the simulator manual.

\subsection{One \acs{eROSITA} telescope}
\label{sec:xml_erosita}

The \ac{XML} code below represents the definition of \texttt{TM1}, one of the
seven \ac{eROSITA} sub-instruments consisting of a telescope and a detector.

{\small \verbatiminput{erosita_1_paper.xml} }

The telescope has a \ac{FOV} with a diameter of $1.02\,\mathrm{deg}$
and a focal length of $1.6\,\mathrm{m}$. The \ac{PSF} and vignetting
function are described in two separate files including several data
sets for different energies and off-axis angles.

The detector consists of $384\times 384$ pixels with an area of
$75\,\mu\mathrm{m}\times 75\,\mu\mathrm{m}$ each. The applied coordinate system
is centered on the detector. Its origin coincides with the intersection of the
optical axis of the telescope with the detector plane and is rotated by
$90^\circ$. The pixels have no borders. The presented detector model does not
include a particular framestore area as the real \ac{eROSITA} \acp{CCD}
\citep[][]{meidinger2011a}. Instead the collected charges in the sensitive
part of the pixel array are read out with the same speed as the transfer time to
the framestore area would require. 

The combined \ac{ARF} of the instrument, which takes into account the effective
area of the telescope and the quantum efficiency of the detector, is stored in
the file \texttt{tm1\_arf\_filter\_000101v02.fits}. The energy resolution of the
detector and related effects such as an escape peak are described by the
\ac{RMF}. A charge transfer efficiency ({CTE)} of $100\,\%$ is assumed and charge cloud splitting between
neighboring pixels is simulated according to an exponential model model
developed by K.~Dennerl (priv.\ comm.). Random background events are inserted
from the PHA background file \texttt{sixte\_ero\_particle.pha}.

The lower threshold for the measuring a signal is set to $50\,\mathrm{eV}$. The
pattern analysis algorithm selects only events with a signal of more than
$150\,\mathrm{eV}$ as main events, which are below an upper threshold of $12\,\mathrm{keV}$.

\begin{figure}
  \includegraphics[width=\columnwidth]{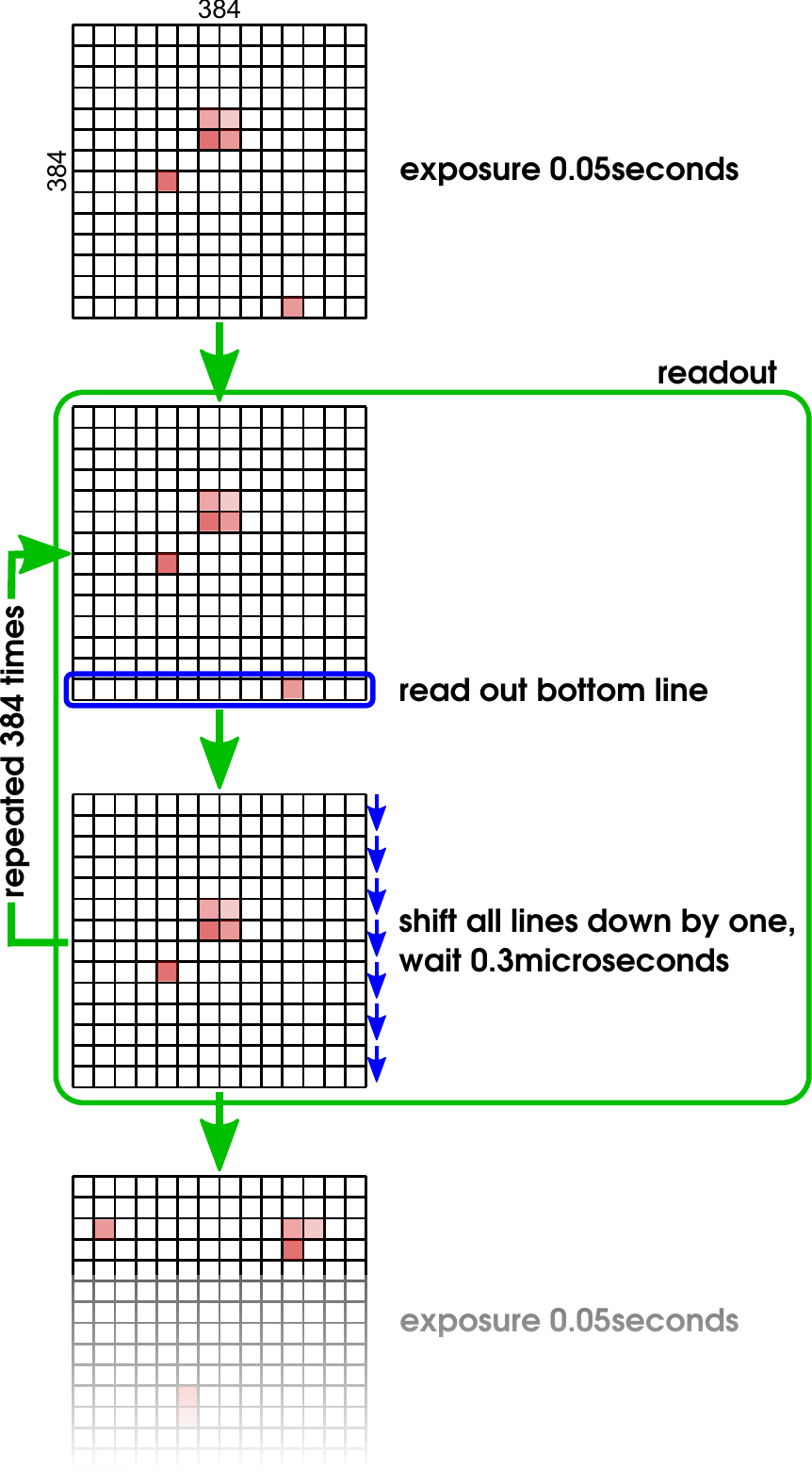}
  \caption[Read out of \acs{eROSITA} \acs{CCD}]{Schematic
    implementation of the read-out mode of an \ac{eROSITA}
    \ac{CCD}. For illustrative purpose not all $384\times 384$ pixels
    are drawn. Each exposure time of $50\,\mathrm{ms}$ is followed by
    $384$ subsequent read-outs and line shifts. As indicated in the
    text, in the simulation the transfer of signals to the framestore
    area, which is typical for the \ac{eROSITA} \acp{CCD}, is replaced
    by a faster read-out with an equivalent time interval.}
  \label{fig:erosita_readout}
\end{figure}

The read-out mode of the \ac{CCD} is characterized by an exposure time
of $50\,\mathrm{ms}$ followed by $384$ subsequent read-outs and line
shifts, which is illustrated in Fig.~\ref{fig:erosita_readout}. As
mentioned above, instead of using a framestore area, for the read-out
process a high line shift rate is used. A single line shift lasts
$0.3\,\mu\mathrm{s}$, that is, the read-out of the whole array takes about
$115\,\mu\mathrm{s}$ and is equivalent to the transfer from the image
to the framestore area.

With this \ac{XML} description the essential parameters of a single
\ac{eROSITA} telescope and detector sub-system are defined. Similar
\ac{XML} definition files for various instruments can be found in the
\ac{SIXTE} software package.

\subsection{\textit{Suzaku} \acs{XIS} burst option}

The \ac{XIS} aboard \textit{Suzaku} in the normal mode provided a
burst option in order to avoid pile-up for observations of bright
X-ray sources. With this option, the exposure time $b$ can be set to
an arbitrary value in $1$/$32$\,s steps \citep{koyama2007a}.

Before the actual exposure takes place, the detector waits for an
interval of $8\,\mathrm{s}-b$. All signals collected in the
$1024\times 1024$ pixels of the \ac{CCD} during this period are then
transferred out of the imaging area and flushed without recording. In
parallel, charge is injected into particular rows of the \ac{CCD}
according to the spaced-row charge injection mechanism, which
mitigates the effects of radiation damage
\citep{bautz2004a,koyama2007a}. The entire procedure results in a
transfer time of $156\,\mathrm{ms}$.

The signals collected during the subsequent science exposure of length
$b$ are transferred to the framestore area and read out from there. As
this second transfer takes place without charge injection, it only
requires $25\,\mathrm{ms}$.

The described detector operation can be implemented in \ac{SIXTE} with the
appropriate \ac{XML} tags, in this case using for an exposure $b=0.094\,\mathrm{s}$:

\verbatiminput{suzaku_burst.xml}

\begin{figure}
  \centering
  \includegraphics[width=0.85\columnwidth]{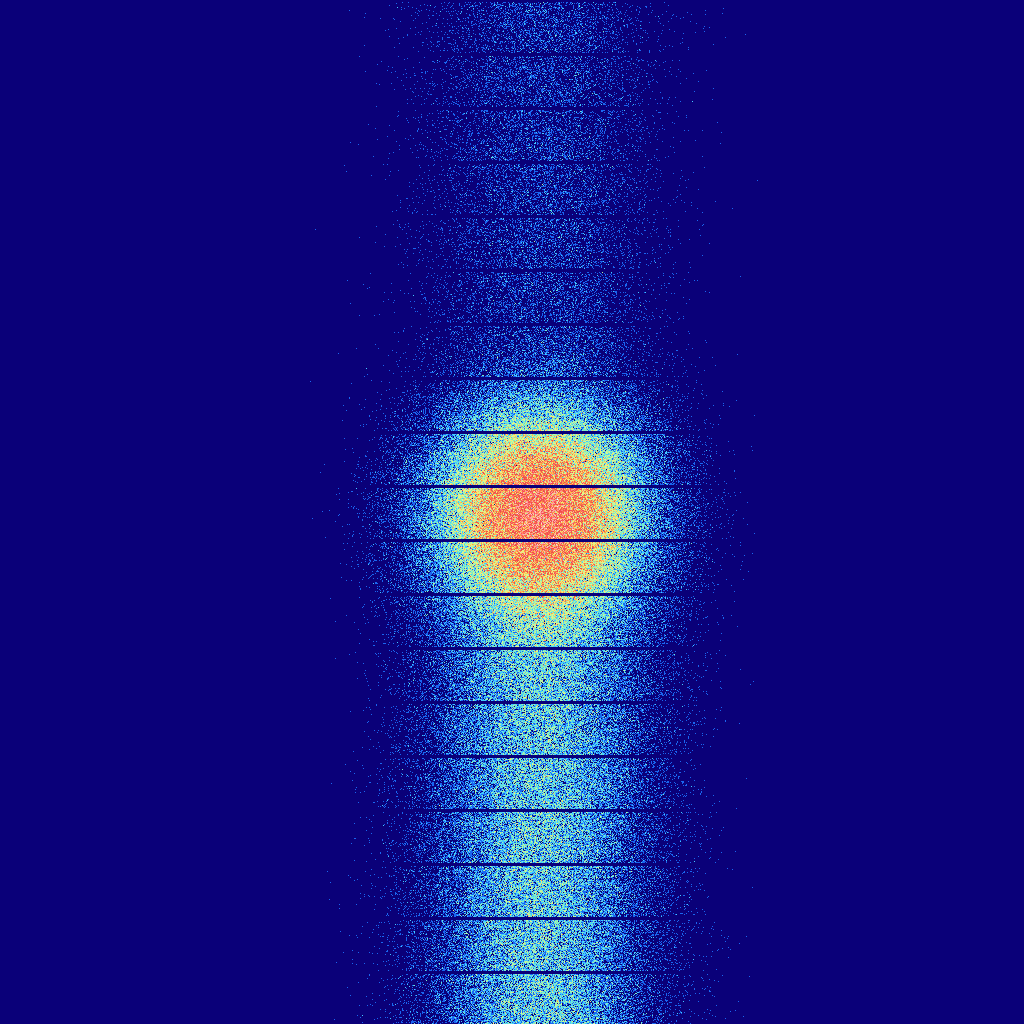}
  \caption[Simulated view of the Crab with \textit{Suzaku} \acs{XIS}
  burst mode]{Simulated observation of the Crab nebula with one of the
    \textit{Suzaku} \acp{XIS} (chip width corresponds to 17~arcmin)
    using the burst option. The read-out direction is towards the
    bottom and the image is given in linear scaling. The evident
    asymmetric density of \ac{OOT} events above and below the source
    is caused by the different speed of the two charge transfer
    processes before and after the science exposure of
    $b=0.094\,\mathrm{s}$.}
  \label{fig:suzaku_burst}
\end{figure}

Analogous to the model of the \ac{eROSITA} \ac{CCD} introduced in Appendix
\ref{sec:xml_erosita}, the framestore area, which is part of the real
device, is not implemented in the detector model, but the charges are
directly read out from the bottom line of the \ac{CCD}. Due to the
different duration of the first and the second charge transfer process
with and without charge injection, the images obtained from
observations in burst mode exhibit an asymmetric distribution of
\ac{OOT}
events\footnote{\url{http://www.astro.isas.jaxa.jp/suzaku/doc/suzaku_td/}},
as illustrated in Fig.~\ref{fig:suzaku_burst}.

\subsection{WFI DEPFET read-out implementation}
\label{sec:depfet}                                                                                                                                                              
                                                                                                                                                                                                        
\begin{figure}                                                                                                                                                                                          
  \centering                                                                                                                                                                                            
    \includegraphics[width=\columnwidth]{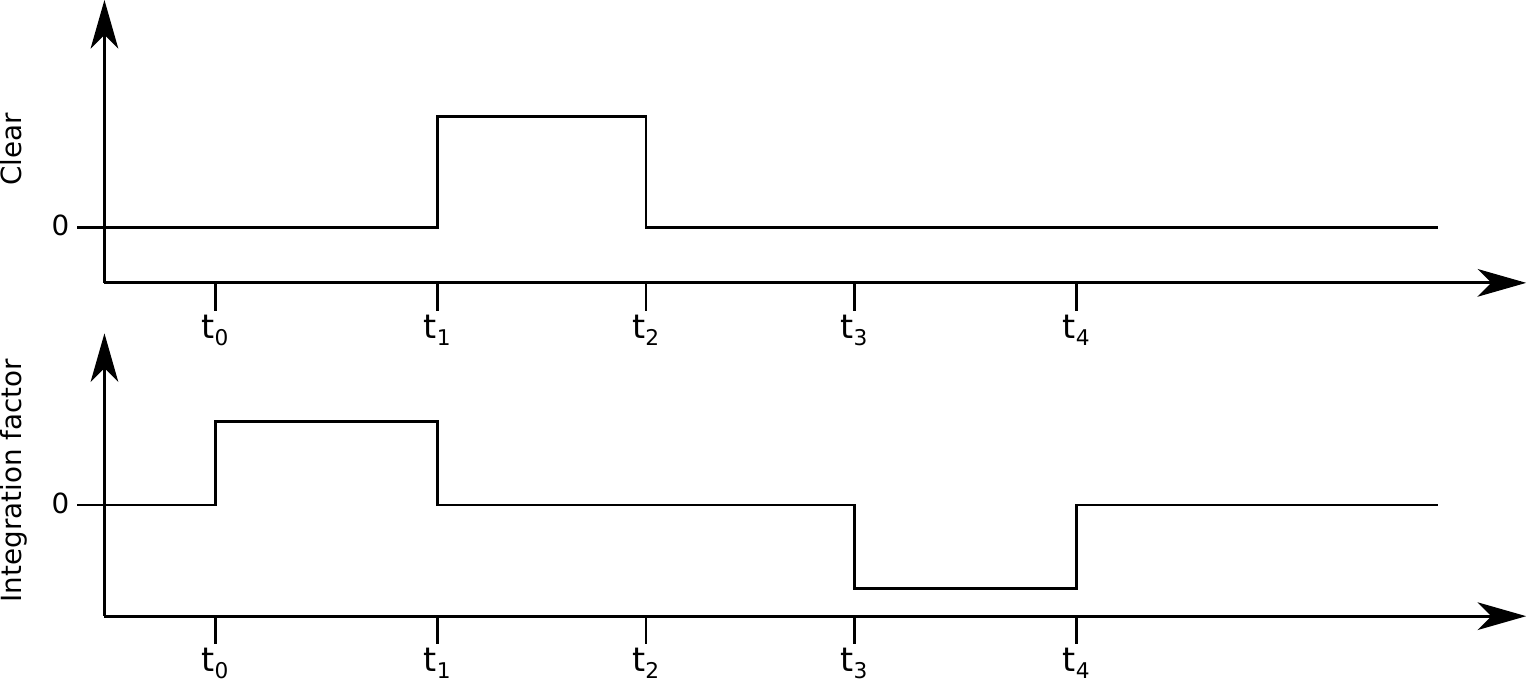}                                                                                                                                                  
  \caption{Basic scheme of how the DEPFET read-out is implemented in \ac{SIXTE}. The read-out 
  is separated in four time steps from $t_0$ to $t_4$, the duration of which can be specified in 
  the instrument description. Between the first and second integration (lower panel), the charge is 
  removed in the clear step (upper panel) followed by a settling time. As displayed in the figure,
  the second integration is performed with a negative sign and therefore acts as
  subtraction of the baseline.  }                                                                                                                                                      
  \label{fig:depfettimes}                                                                                                                                                                               
\end{figure}

This section describes the implementation of the WFI DEPFET. The DEPFET
read-out is not an instantaneous determination of the energy of the photon but it
integrates first the voltage signal of the photon together with the signal baseline,
and after the charge of the photon is removed by the clear, the baseline is
integrated again to remove it from the measurement. In a simplified scheme, the
measurement can be described with three time intervals and the corresponding
integration factors, as depicted in Fig.~\ref{fig:depfettimes}.
                                                                                                                                                                                                        
                                                                                                                                                                                                        
After the initial settling, the read-out of a DEPFET begins at $t_0$. Between
$t_0$ and $t_1$, the signal present at the internal gate is integrated. Between
$t_1$ and $t_2$, any charge is removed from the internal gate of the DEPFET .  Between
$t_2$ and $t_3$, the second settling occurs.  Between $t_3$ and $t_4$, the
signal is integrated again but with negative sign. The result is that the
baseline signal can be subtracted from the first measurement, providing an
estimate of the signal of the photon.
                                                                                                                                                                                                        
Misfits are photons which hit the detector during the integration time
intervals. As their signal is not integrated for the full integration time, they
can produce incorrect measurements. The resulting measurement $E_m$ of their true
energy $E_p$ can be described as dependent on their impact time $t_p$:
                                                                                                                                                                                                        
\begin{itemize}                                                                                                                                                                                         
  \item $t_p<t_0$ or $t_p>t_4$: $E_m=E_p$                                                                                                                                                               
  \item $t_0<t_p<t_1$: $E_m=E_p\times (t_1-t_p)/(t_1-t_0)$                                                                                                                                              
  \item $t_2<t_p<t_3$: $E_m=-E_p$                                                                                                                                                                       
  \item $t_3<t_p<t_4$: $E_m=-E_p\times (t_4-t_p)/(t_4-t_3).$                                                                                                                                             
\end{itemize}                                                                                                                                                                                           
                                                                                                                                                                                                        
In the case of $t_3<t_p<t_4$, the charge is not removed in this read-out cycle
and will be measured again during the next cycle, before it is cleared.
                                                                                                                                                                                                        
Another effect to be simulated is the limited clear speed. Instead of removing
the charge instantaneously, it is cleared linearly between $t_1$ and $t_2$. If a
photon hits the detector during this interval and causes a charge $q_p$, it is
only partially cleared and leaves a remaining charge $q_r$, according to
\begin{displaymath}
        q_r=q_p\times (t_p-t_1)/(t_2-t_1) \,\,.                                                                                                                                                         
\end{displaymath}
The remaining charge will be measured in the second integration time and causes a negative energy measurement of the value                                                                              
\begin{displaymath}                                                                                                                                                                                     
        E_m=-E_p \times (t_p-t_1)/(t_2-t_1) \,\,.                                                                                                                                                       
\end{displaymath}                                                                                                                                                                                       
The charge will be measured again in the next read-out cycle. If the 
impact is between $t_2$ and $t_3$, the charge is remaining completely.                                                                                                                                  
If two photons impact in the same pixel during one read-out cycle,
their signals add to each other.

The parameters defining the DEPFET read-out can easily be set in the XML file. In case of the
full-frame read-out of on large detector of the WFI, the relevant lines are: 
{\small \verbatiminput{ld_wfi_ff_large.xml} }

\section{Charge cloud models}
\label{sec:charge_cloud}

As indicated in Sect.~\ref{sec:detector_model}, \ac{SIXTE} needs a model to
simulate the distribution of charge between neighboring pixels. The standard
``Gaussian'' model assumes a 2D, rotational symmetric charge cloud.
We note that \ac{SIXTE} also implements a specific charge cloud model,
called ``Exponential'', tailored specifically for \ac{eROSITA} (K.~Dennerl,
priv.\ comm.). 

\begin{figure}
  \centering
  \includegraphics[width=\columnwidth]{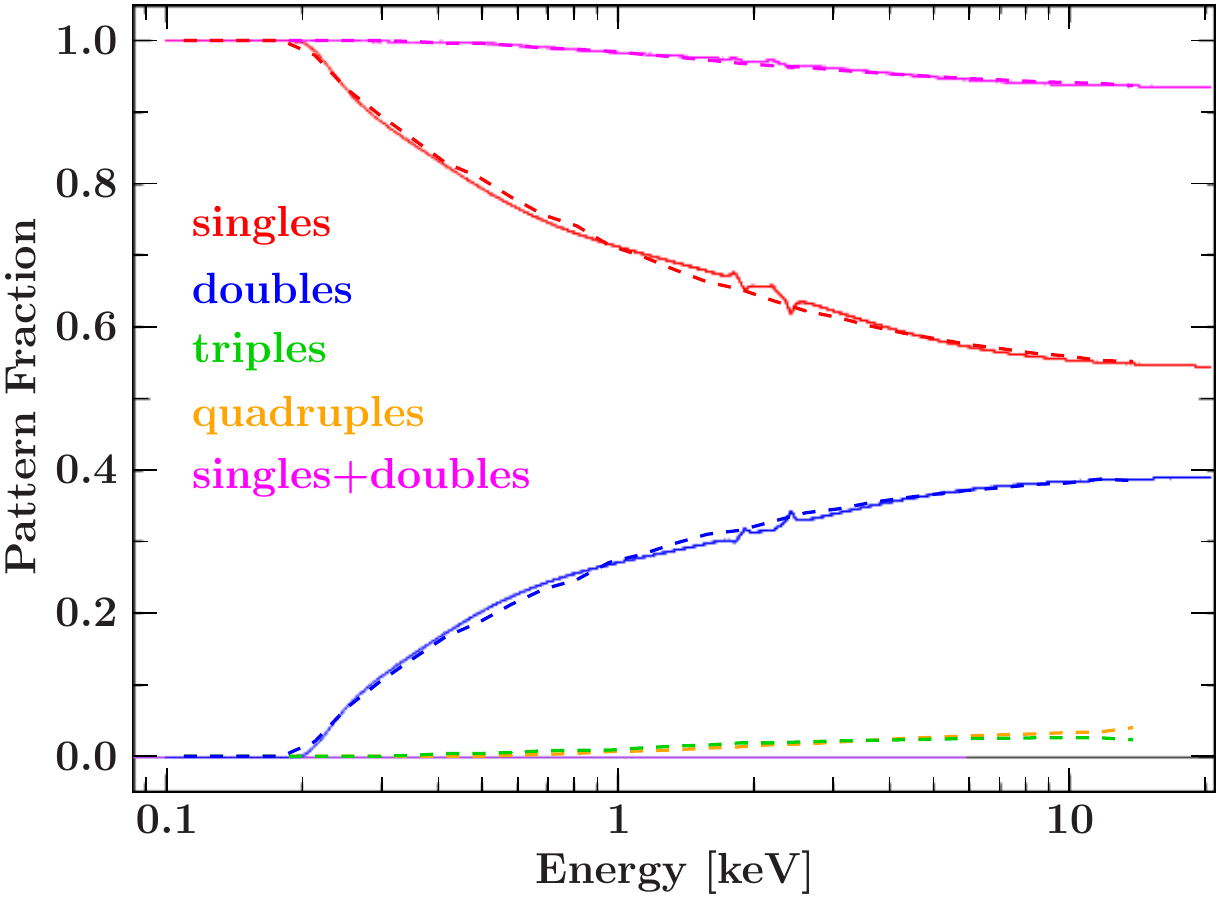}
  \caption[Comparison of Gaussian charge cloud model with
  \texttt{epatplot}]{Comparison of the Gaussian charge cloud model
    (dashed lines) with the distribution of pattern types obtained
    with the SAS task \texttt{epatplot} (solid lines) for an observation
    of Cen A with the
    \ac{EPIC}~pn camera on \ac{XMM} operated in Small Window.  The
    simulated distribution matches very well the values expected from
    the instrument model.}
  \label{fig:epatplot}
\end{figure}

In the Gaussian model, the charge fraction collected in the detector pixel with
the corners $(x_n,y_m)$, $(x_{n+1},y_m)$, $(x_{n+1},y_{m+1})$, and
$(x_n,y_{m+1})$ can be determined as \citep{popp2000a}:
\begin{equation}
  \label{equ:gaussian_charge_fraction}
  c_{n,m}=\frac{c_\mathrm{total}}{2\pi\sigma^2}\cdot \int\limits_{x_n}^{x_{n+1}} \mathrm{e}^{-\frac{(x-x_\mathrm{i})^2}{2\sigma^2}}\,\mathrm{d}x \int\limits_{y_m}^{y_{m+1}} \mathrm{e}^{-\frac{(y-y_\mathrm{i})^2}{2\sigma^2}}\,\mathrm{d}y
,\end{equation}
where $(x_\mathrm{i},y_\mathrm{i})$ denotes the impact position of the
photon and $c_\mathrm{total}$ the total generated charge. As the
charge cloud size ($\sigma\sim 10\,\mu\mathrm{m}$) is typically
smaller than the size of the detector pixels,
Eq.~(\ref{equ:gaussian_charge_fraction}) is evaluated only for the
four pixels directly surrounding the impact position. By
adjusting the size of the charge cloud and the lower detection
threshold for split partners, the simulated pattern distribution can
be adjusted to fit measurements with real detectors, as shown in
Fig.~\ref{fig:epatplot} for the \ac{EPIC}~pn camera on \ac{XMM}.

\end{document}